\documentclass[acmtog]{acmart}

\AtBeginDocument{%
  \providecommand\BibTeX{{%
    \normalfont B\kern-0.5em{\scshape i\kern-0.25em b}\kern-0.8em\TeX}}}

\acmJournal{TOG}
\setcopyright{acmlicensed}
\acmYear{2024} 
\acmVolume{43} 
\acmNumber{4} 
\acmArticle{39} 
\acmMonth{7}
\acmDOI{10.1145/3658170}

\usepackage{enumitem}

\newcommand{\revision}[1]{{\color{black} {#1}}}

\usepackage{color}
\definecolor{applegreen}{rgb}{0.55, 0.71, 0.0}
\definecolor{autumnorange}{rgb}{0.87, 0.61, 0.33}

\begin{document}

\title{DreamMat: High-quality PBR Material Generation with Geometry- and Light-aware Diffusion Models
}

\author{Yuqing Zhang}
\authornote{Equal contribution}
\affiliation{
  \institution{State Key Lab of CAD\&CG, Zhejiang University}
  \city{Hangzhou}
  \state{Zhejiang}
  \country{China}
}
\email{3180102110@zju.edu.cn}

\author{Yuan Liu}
\authornotemark[1]
\affiliation{%
  \institution{Tencent Games}
  \city{Shenzhen}
  \country{China}
  }
\email{liuyuanwhuer@gmail.com}

\author{Zhiyu Xie}
\affiliation{
  \institution{State Key Lab of CAD\&CG, Zhejiang University}
  \city{Hangzhou}
  \state{Zhejiang}
  \country{China}
}
\email{xiezhiyu@zju.edu.cn}

\author{Lei Yang}
\affiliation{%
  \institution{Tencent Games}
  \city{Shenzhen}
  \country{China}
  }
\email{vfxylapply@gmail.com}
  
\author{Zhongyuan Liu}
\affiliation{%
  \institution{Tencent Games}
  \city{Shenzhen}
  \country{China}
  }
\email{zyliu28@mail.ustc.edu.cn}

\author{Mengzhou Yang}
\affiliation{%
  \institution{Tencent Games}
  \city{Shenzhen}
  \country{China}
  }
\email{ yangskin@163.com}

\author{Runze Zhang}
\affiliation{%
  \institution{Tencent Games}
  \city{Shenzhen}
  \country{China}
  }
\email{ryanrzzhang@tencent.com}
  
\author{Qilong Kou}
\affiliation{%
  \institution{Tencent Games}
  \city{Shenzhen}
  \country{China}
  }
\email{kouqilong1988@gmail.com}

\author{Cheng Lin}
\affiliation{%
  \institution{Tencent Games}
  \city{Shenzhen}
  \country{China}
  }
\email{chlin@connect.hku.hk}

\author{Wenping Wang}
\affiliation{%
  \institution{Texas A\&M University}
  \city{Texas}
  \country{U.S.A}
  }
\email{wenping@tamu.edu}

\author{Xiaogang Jin}
\authornote{Corresponding author.}
\affiliation{
  \institution{State Key Lab of CAD\&CG, Zhejiang University}
  \city{Hangzhou}
  \state{Zhejiang}
  \country{China}
}
\email{jin@cad.zju.edu.cn}








\begin{abstract}
Recent advancements in 2D diffusion models allow appearance generation on untextured raw meshes. These methods create RGB textures by distilling a 2D diffusion model, which often contains unwanted baked-in shading effects and results in unrealistic rendering effects
in the downstream applications. 
Generating Physically Based Rendering (PBR) materials instead of just RGB textures would be a promising solution. However, directly distilling the PBR material parameters from 2D diffusion models still suffers from incorrect material decomposition, such as baked-in shading effects in albedo. 
We introduce \textit{DreamMat}, an innovative approach to resolve the aforementioned problem, to generate high-quality PBR materials from text descriptions.
We find out that the main reason for the incorrect material distillation is that large-scale 2D diffusion models are only trained to generate final shading colors, resulting in insufficient constraints on material decomposition during distillation.
To tackle this problem, we first finetune a new light-aware 2D diffusion model to condition on a given lighting environment and generate the shading results on this specific lighting condition. 
Then, by applying the same environment lights in the material distillation, DreamMat can generate high-quality PBR materials that are not only consistent with the given geometry but also free from any baked-in shading effects in albedo.
Extensive experiments demonstrate that the materials produced through our methods exhibit greater visual appeal to users and achieve significantly superior rendering quality compared to baseline methods, which are preferable for downstream tasks such as game and film production. Project page: \url{https://zzzyuqing.github.io/dreammat.github.io/}.

\end{abstract}


\begin{teaserfigure}
\centering
  \includegraphics[width=0.95\textwidth]{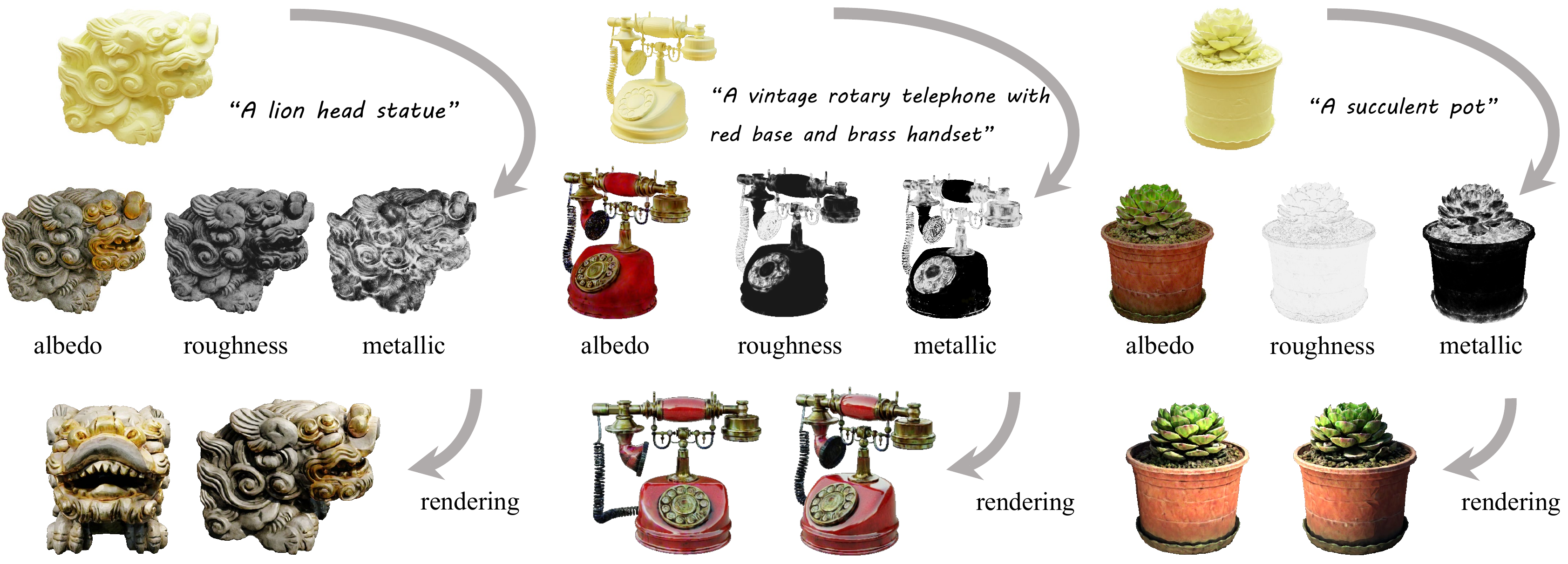}
  \caption{Using untextured meshes and textual descriptions as input (top row), our method generates high-quality \revision{appearances consisting of} albedo, roughness, and metallic (middle row) that can be applied in modern graphics engines for photo-realistic rendering under any new illumination environments (bottom row).}
  \label{fig:teaser}
\end{teaserfigure}

\maketitle
\section{Introduction}
\begin{figure}
  \includegraphics[width=\linewidth]{{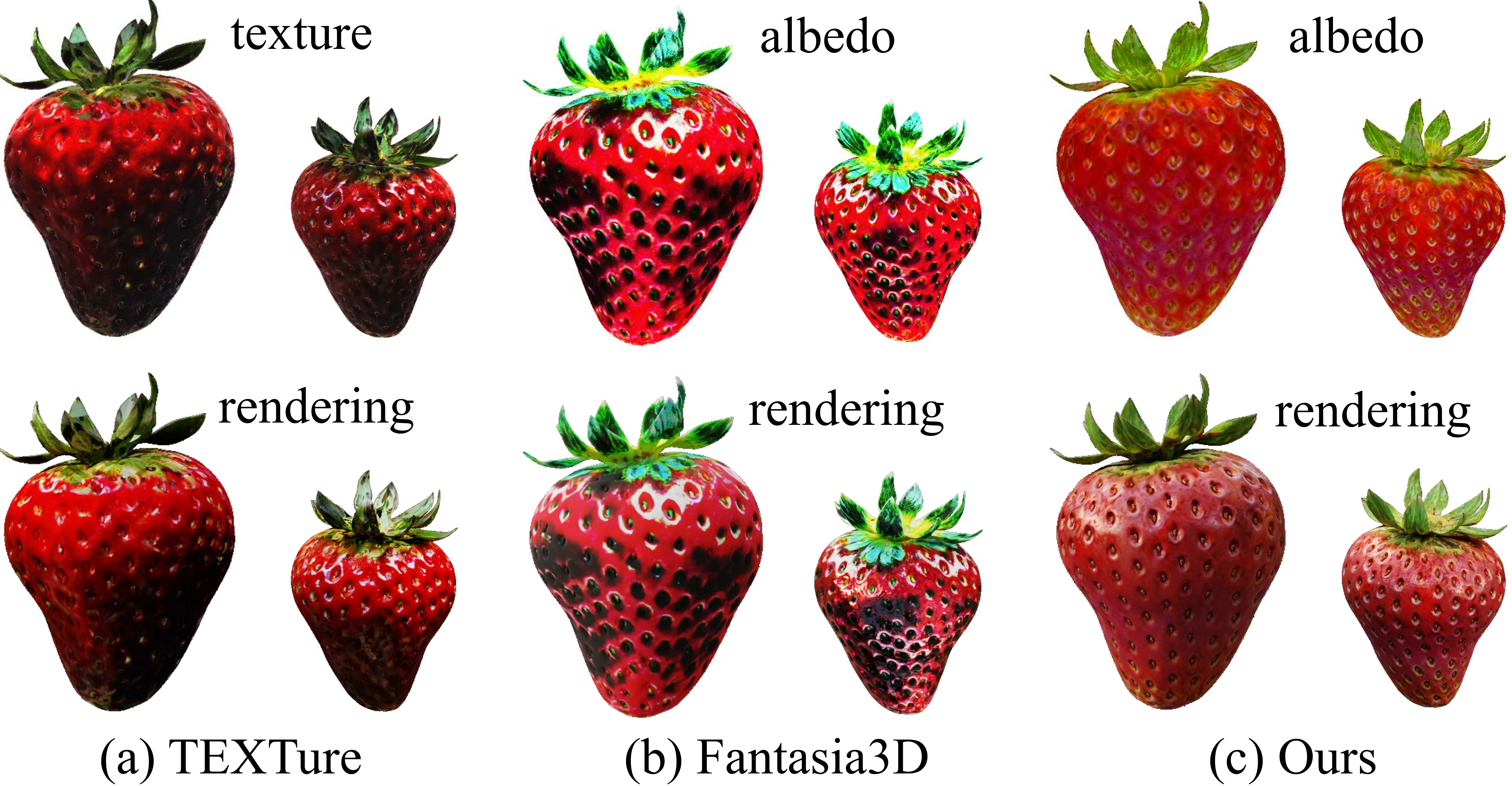}}
  \caption{ \textbf{Generated albedo and rendering results in the same environment light.} (a) TEXTure~\cite{yu2023texture} generates an RGB texture map containing shading effects, leading to incorrect
  renderings in a new environment. (b) Fantasia3D~\cite{Chen_2023_ICCV} directly distills a diffusion model to generate materials, which still contain unwanted shading effects in albedo. (c) Our method can generate correct materials, allowing for more photorealistic renderings in a new environment.}
  \label{fig:light}
  \vspace{1mm}
\end{figure}

Creating high-quality appearances for objects is a critical task in computer graphics because they can significantly improve the realism of rendering in a variety of applications such as movies, games, and AR/VR. However, even experienced artists may find it time-consuming to create object appearances \cite{labschutz2011content} due to the need for expertise with complex commercial 3D software such as Mari, ZBrush, and Substance Painter. This necessitates the development of new tools for efficiently creating object appearances, such that even novice users can do so with simple text prompts.
Several text-driven methods \cite{richardson2023texture,chen2023text2tex,cao2023texfusion, yu2023texture, le2023euclidreamer, knodt2023consistent, zeng2023paint3d} have been developed to generate RGB textures on untextured meshes. These approaches use powerful 2D text-to-image diffusion models \cite{rombach2022high,ho2020denoising}, such as the Stable Diffusion model \cite{rombach2022high}, to achieve impressive results. While these techniques make it easier to create object appearances, they frequently produce undesirable shading effects such as highlights and shadows, as shown in Fig. \ref{fig:light} (a).
The baked-in shading effects in generated appearances cause unrealistic results in rendering, limiting their applicability in downstream tasks such as game or film production.

Using PBR materials instead of RGB textures can improve object appearance, but it may present some challenges. Directly training a material generation network is difficult and costly due to a scarcity of high-quality 3D assets with known PBR materials. 
Alternatively, recent research~\cite{Chen_2023_ICCV,xu2023matlaber} integrates material decomposition in distilling a powerful 2D text-to-image diffusion model, which does not require training on datasets with known PBR materials. Though these methods achieve impressive results in some examples, correct material generation remains a challenge, as illustrated in Fig.~\ref{fig:light} (b). This is because 2D diffusion generative models can only generate final shading results, making it difficult to decompose correct material parameters due to the ill-posed nature of the material decomposition task.

\begin{figure}
  \includegraphics[width=\linewidth]{{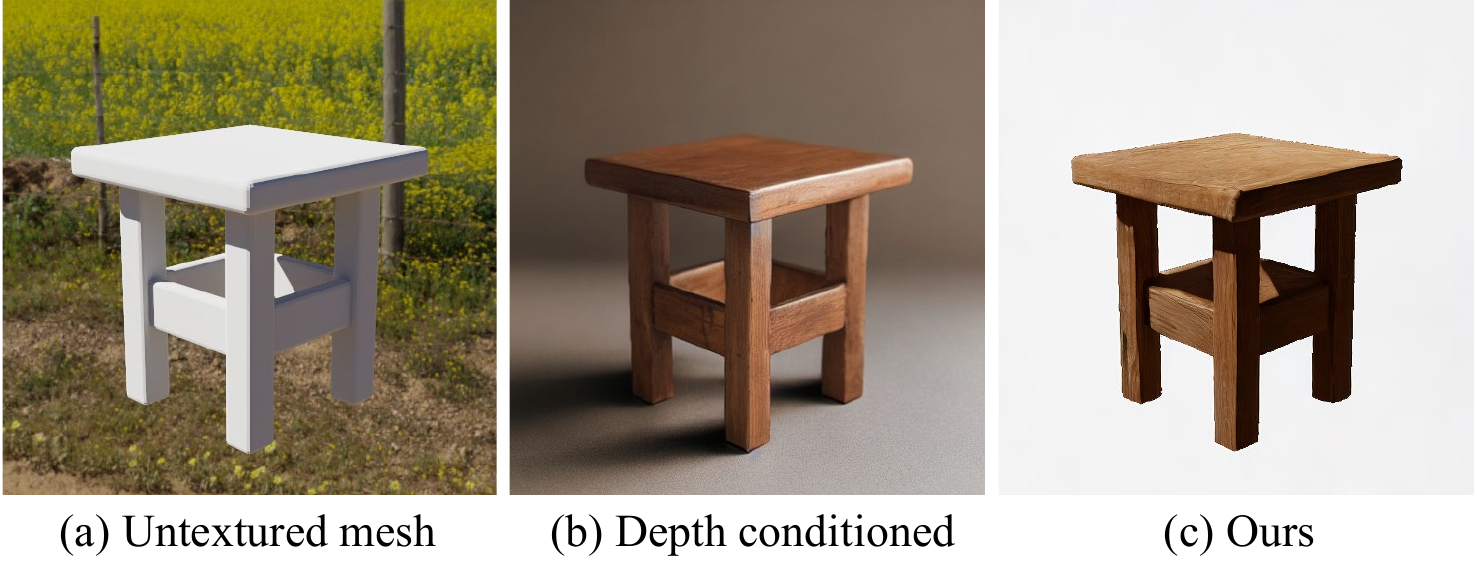}}
  \caption{\textbf{An untextured stool mesh and its generated images using different methods.} 
  (a) An untextured mesh with a given light environment. (b) A generated image of a depth-to-image Stable Diffusion model, which is inconsistent with the given environment light and results in incorrect materials decomposition. (c) An image generated by our geometry- and light-aware diffusion model, which is consistent with the environment light.}
  \label{fig:light_aware}
  
\end{figure}

In this paper, we begin with an in-depth analysis of the ill-posed problem of material decomposition in the context of diffusion distillation framework. \revision{Diffusion models are trained to generate natural RGB images which are the final shading results of some unknown environmental lights and materials.}
However, on different distillation steps, the generated images may correspond to different environmental lights, making it impossible to accurately estimate a fixing environmental light and thus leading to incorrect materials.
Fantasia3D~\cite{Chen_2023_ICCV} uses a single predefined environmental light to distill diffusion models to generate materials but the generated images from diffusion models may not be consistent with the given environment light, as shown in Fig.~\ref{fig:light_aware} (b), still resulting in incorrect materials. 

Based on our analysis, we present \textit{DreamMat}, \revision{a novel method to create high-quality appearances on an untextured mesh by generating PBR materials with a diffusion model.} The key idea of DreamMat consists of two aspects. First, in the distillation process, we randomly select from a set of predefined HDR images as the environment light so that DreamMat can focus on generating the object materials. Second, we propose a novel geometry- and light-aware diffusion model, which is trained to generate images that are consistent with the given environment light, as shown in Fig.~\ref{fig:light_aware} (c). By distilling this new geometry- and light-aware diffusion, DreamMat is able to accurately \revision{generate materials, enabling rendering photo-realistic images in various environments, outperforming baseline methods by a significant margin, and also being more compatible with modern graphics engines as shown in Fig.~\ref{fig:light} (c)}.

Our contributions are summarized as follows:
\begin{itemize}[noitemsep,nolistsep,leftmargin=*]
    \item A geometry- and light-aware diffusion model to generate images consistent with geometric and light contexts.
    \item A novel framework for text-guided material generation on specified meshes with high-quality albedo, roughness, and metallic.
\end{itemize}
\vspace{-10pt}

\section{Related Work}

\subsection{BRDF estimation}
\par 
Surface BRDF estimation from images relies on inverse rendering techniques \cite{barron2014shape, mitsuba2019}.
Many studies infer the geometry and material properties of real-world objects from image collections under controlled light conditions \cite{bi2020deep, nam2018practicle, xia2016recovering} or domain-specific priors \cite{wimbauer2022rendering, li2018learning, li2020inverse, barron2014shape, ye2023intrinsicnerf, guo2020materialgan, duan2019deep} to reduce ambiguities in the inverse rendering.
With the rise of neural rendering exemplified by NeRF~\cite{mildenhall2021nerf}, inverse rendering has also made significant qualitative progress. To model spatially-varying bidirectional reflectance distribution function (SVBRDF) under more casual capture conditions, many methods \cite{boss2021nerd, boss2021neural, zhang2022iron, yariv2020multiview, physg2021, Munkberg_2022_CVPR, dave2022pandora, sun2023neural, ye2023intrinsicnerf, cheng2024structure} have relied on implicit representation to provide geometry prior. Subsequent works improve the quality of the reconstruction results by introducing visibility prediction~\cite{srinivasan2021nerv, chen2022tracing}, modeling indirect illumination~\cite{zhang2022invrender, Jin2023TensoIR, yao2022neilf, zhang2023neilf++, deng2022dip,  yang2023sire}, applying  Monte Carlo sampling~\cite{luan2021unified, hasselgren2022shape,zhu2023i2, li2023neisf, liu2023nero, tg2023neural}, \revision{utilizing deep polarization information~\cite{zhao2022polarimetric, deschaintre2021deep}, } setting multiple flashlights~\cite{bi2020neural,kuang2022NeROIC,cheng2021multi,li2022neural, yang2022psnerf}, and introducing material priors~\cite{zhang2021nerfactor, boss2021neural}. 

Recent works~\cite{jiang2023gaussianshader, gao2023relightable, liang2023gs} apply 3D Gaussian Splatting~\cite{kerbl3Dgaussians} to inverse rendering, achieving speed improvements. Other works~\cite{kocsis2023intrinsic, lyu2023diffusion} integrate a diffusion model into the traditional inverse rendering framework. Their improved decomposition results can be attributed to the strong learned prior of diffusion models trained on large-scale real-world images.

Contrary to the aforementioned works that perform inverse rendering based on ground truth images, our endeavor is not one of reconstruction but generation. In our framework, both geometry and lighting conditions are predefined, facilitating the generation of materials through a novel application of material decomposition integrated with a 2D diffusion model. Our method builds upon the second stage of NeRO \cite{liu2023nero} and employs a simplified Disney BRDF \cite{burley2012physically} to regulate material parameters across various established lighting scenarios.
 \begin{figure*}
  \includegraphics[width=0.95\linewidth]{{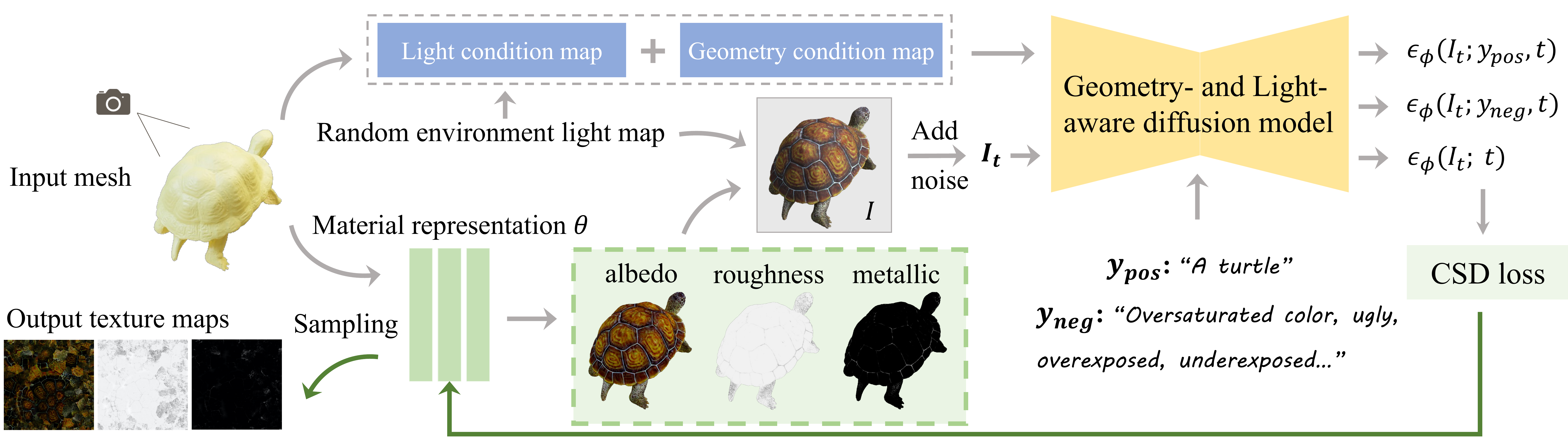}}
  \caption{\revision{\textbf{Overview of our pipeline.} DreamMat distills a diffusion model to generate PBR materials.
  We first use Monte Carlo sampling to render images of the object from its material representation and a randomly-selected predefined environment light. Then, we train the material representation by CSD loss on rendered images using a geometry- and light-aware diffusion model.}}
  \label{fig:pipeline}
  \vspace{-2mm}
\end{figure*}
\subsection{Text-guided 3D appearance generation}
Recent advances in large language models and image diffusion techniques demonstrate their remarkable capability for text-to-3D generation. The pioneering work Dreamfusion \cite{poole2022dreamfusion} first proposes the SDS loss to distill 3D assets from pre-trained text-to-image diffusion models. This idea inspires a series of following works to extend text-to-3D generation to material \cite{Chen_2023_ICCV, xu2023matlaber, youwang2023paint, liu2023unidream}, image-to-3D \cite{metzer2022latent, liu2023zero, qian2023magic123, liu2023one2345, tewari2023forwarddiffusion, zhou2023sparsefusion}, and higher quality text-to-3D \cite{lin2023magic3d, wang2023prolificdreamer, sun2023dreamcraft3d, sweetdreamer, katzir2023noise, yu2023csd, zhu2023hifa} generation.

A notable trend in this field is the texture generation.  TEXTure \cite{richardson2023texture} and Text2Tex \cite{chen2023text2tex} use an iteratively inpainting method for texturing the given mesh with depth-conditioned Stable Diffusion. However, they exhibit noticeable artifacts at the viewpoint junctions. Subsequent works \cite{cao2023texfusion, yu2023texture, knodt2023consistent, Tang2023mvdiffusion, zhang2023repaint123,liu2023text, oh2023controldreamer,tang2023text,cheskidova2023geometry} improve the 3D consistency and quality of the generated textures. Many works on generating textures for human~\cite{Svitov_2023_ICCV,Ma_2023_ICCV,albahar2023single,huang2023humannorm} or rooms~\cite{chen2023scenetex, hollein2023text2room,yang2023dreamspace, wen2023anyhome} have also been proposed. 
However, the generated RGB texture maps often contain baked-in highlights or shadows, disabling realistic rendering in downstream tasks.

\par To address this limitation, recent works attempt to incorporate BRDF to generate more realistic appearances. \revision{Previous Methods~\cite{zhou2022tilegen, MatFusion_2023,vecchio2023matfuse,vecchio2023controlmat}
can generate PBR material maps given the input image. However, these methods primarily generate materials in the 2D image space and cannot generate appearances for 3D meshes. }
TANGO~\cite{chen2022tango} employs CLIP~\cite{radford2021learning} loss to generate both lighting and material properties simultaneously for a given mesh. However, this method produces materials with less detail and the use of position encoding can lead to grid-like artifacts. Fantasia3D~\cite{Chen_2023_ICCV} integrates material decomposition into a distillation-based text-to-3D framework but faces challenges with baked-in shading effects in albedo. There are some concurrent works Paint3D~\cite{zeng2023paint3d}, UniDream~\cite{liu2023unidream} which train diffusion models to generate lightless albedo. In comparison, our method does not require ground-truth albedo data for training and is based on distillation loss. The concurrent work Matlaber~\cite{xu2023matlaber} introduces a material prior distribution to address the ill-posed problem while our method relies on the fixed environmental light and light-aware diffusion model. Another concurrent work Paint-it~\cite{youwang2023paint} is similar to Fantasia3D but with CNN-parameterized materials.

\section{Method}
\subsection{Overview}
Given an untextured mesh and a textual prompt, the target of our method is to generate high-quality Spatially Varying Bidirectional Reflectance Distribution Function (SVBRDF) materials on the mesh according to the text prompt. 
An overview of our method is illustrated in Fig.~\ref{fig:pipeline}. 
First, to represent the SVBRDF material, we adopt the hash-grid-based representation to store the SVBRDF parameters and apply the rendering equation to render images of the given mesh from this representation (see Sec.~\ref{sec:inv}). Then, as introduced in Sec.~\ref{sec:csd}, we add noises to the rendered images and apply a diffusion model to denoise the image, which results in a distillation loss to learn the parameters in the hash-grid representation. To avoid the baked-in lighting and shadows, we adopt a geometry- and light-aware diffusion model as the distillation diffusion model as stated in Sec.~\ref{sec:diffusion}. Finally, the generated materials can be exported to contract material maps defined on the provided or extracted UV map of the mesh, which can be used in editing or arbitrary modern graphics engines.

\subsection{\revision{Material Representation}}
\label{sec:inv}
In this section, we introduce our material representation and how to render images from these representations. We use the hash-grid-based representation from Instant-NGP~\cite{muller2022instant}  to represent the simplified Disney BRDF~\cite{burley2012physically}. The BRDF parameters on a point $\mathbf{p}$, including \revision{albedo $\mathbf{c}$, roughness $\alpha$, and metalness $m$}, are computed by:
\begin{equation}
   (\mathbf{c}, \alpha, m) = \Gamma_\theta(\mathbf{p}),
\end{equation}
where $\Gamma_\theta$ means the hash-grid-based representation with corresponding material parameters $\theta$. Our goal is to learn the trainable parameters $\theta$ so that we can compute the BRDF parameters for any given point $\mathbf{p}$ on the object surface.

\revision{
Following the rendering equation~\cite{render_equation}, the rendering color $L(\mathbf{p},\mathbf{\omega_o})$ for the point $\mathbf{p}$ on the direction $\mathbf{\omega}_o$  is:
\begin{equation}
L(\mathbf{p},\mathbf{\omega_o})=\int_{\Omega}L_i(\mathbf{\omega_i})f(\mathbf{\omega_i},\mathbf{\omega_o})(\mathbf{\omega_i} \cdot \mathbf{n})d\omega_i,
\end{equation}
where $L(\mathbf{\omega_i})$ is the input environmental light, $\mathbf{n}$ is the normal direction, and the BRDF $f(\mathbf{\omega_i},\mathbf{\omega_o})$ is a Cook-Torrance microfacet specular shading model~\cite{cook1982reflectance} which is defined as:

\begin{equation}
f(\mathbf{\omega_i},\mathbf{\omega_o})=\frac{D\ F\ G}{4(\mathbf{\omega_o} \cdot \mathbf{n})(\mathbf{\omega_i} \cdot \mathbf{n})},
\end{equation}
where $D$, $G$ and $F$ are functions representing the GGX~\cite{cook1982reflectance},
the normal distribution function (NDF), geometric attenuation and Fresnel term, respectively.

Following the methodology proposed by~\cite{karis2013real}, we adopt an importance-based Monte Carlo (MC) sampling strategy to separate the rendering equation into diffuse and specular components:
\begin{equation}
L(\mathbf{p},\mathbf{\omega_o})=L_\text{diffuse}+L_\text{specular},
\end{equation}
\begin{equation}
    L_{\text{diffuse}} = \frac{\mathbf{c}}{N_d} \sum_{i=1}^{N_d}   L(\omega_i),
\end{equation}
\begin{equation}
    L_{\text{specular}} = \frac{1}{N_s} \sum_{i=1}^{N_s}  \frac{F(\mathbf{c},m)G(\omega_o,\omega_i,\mathbf{n},\alpha)(\omega_o \cdot \mathbf{h})}{ (\mathbf{n} \cdot \mathbf{h})  (\mathbf{n} \cdot \omega_o)}  L(\omega_i),
\end{equation}
where $N_d$ and $N_s$ denote the number of samples for diffuse and specular components, $\mathbf{h}=(\omega_i+\omega_o)/|\omega_i+\omega_o|$ is the half-way vector. The diffuse component is evaluated using a cosine-weighted hemisphere sampling, while the specular component employs sampling based on the GGX distribution.
}

In an inverse rendering pipeline~\cite{zhang2023neilf++}, both the environmental lights $L(\omega_i)$ and the material parameters $\mathbf{\theta}$ should be estimated. Since we only want to generate material parameters for the given object, we fix the environment lights by randomly selecting a known HDR image as the environment light. This makes the inverse rendering problem less ill-posed for better material generation.
\begin{figure*}
  \includegraphics[width=0.95\linewidth]{{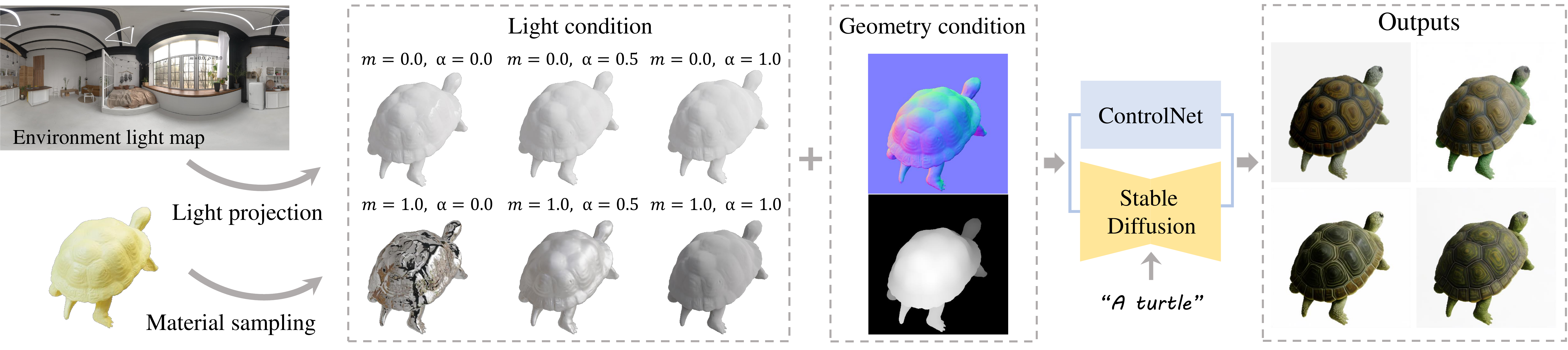}}
  \caption{\revision{\textbf{Our geometry- and light-aware diffusion model} uses an object's normal and depth maps as geometry conditions and six predefined materials with a given environment light as lighting conditions. Our model generates images that align with the given geometry and environment light. }}
  \label{fig:controlnet}
  \vspace{-2mm}
\end{figure*}

\subsection{\revision{Distillation Loss for Material Generation}}
\label{sec:csd}
\revision{The material representation is randomly initialized at first and we follow the Score-Distillation Sampling (SDS) loss~\cite{poole2022dreamfusion} to distill a text-to-image diffusion model to learn the parameters of the material representation. Given a rendered image $I$ from the current material representation, we add a noise $\epsilon_t$ to it to get a noisy image $I_t=I+\epsilon_t$. Then the noisy image is denoised by the diffusion model to get a denoised image $I'_t$ conditioned on the text prompt $y_{\text{pos}}$ and the difference between the denoised image and the rendered image $\delta(I_t)=I'_t-I$ is computed as the distillation loss $\mathcal L_{\text{Distill}}$ and we use the following gradient to optimize $\theta$:
\begin{equation}
    \label{equ:csd}
    \nabla_{\mathbf{\theta}} \mathcal L_{\text{Distill}} = \mathbb E_{t} \left[ \delta(I_t) \frac{\partial I}{\partial \mathbf{\theta}} \right].
\end{equation}
By denoising $I_t$ to $I'_t$ using the diffusion model, $I'_t$ will be more consistent with the given text prompt $y_{\text{pos}}$ than $I$. Thus, minimizing the difference $\delta(I_t)=I'_t-I$ makes the rendered image more aligned with the given text prompts. Thus, Eq.~\eqref{equ:csd} gradually optimizes the material parameters $\theta$ to make the renderings consistent with the input texts.

Here, we adopt a variant of the original SDS loss~\cite{poole2022dreamfusion} called Classifier Score Distillation (CSD) loss~\cite{yu2023csd}, which shows better distillation performance than the SDS loss as demonstrated in Fig.~\ref{fig:sds_csd}.
CSD loss is also defined by Eq.~(\ref{equ:csd}) but utilizes all the null, positive, and negative prompts to compute $\delta{(I_t)}$ as follows:
\begin{equation}
\label{equ:csd-loss}
    \delta(I_t)=\eta_1 \epsilon_{\phi}(I_t; y_{\text{pos}}, t) + (\eta_2 - \eta_1) \epsilon_{\phi}(I_t; t) - \eta_2 \epsilon_{\phi}(I_t; y_{\text{neg}}, t),
\end{equation}
where $t$ is a sampled time step, $\epsilon_{\phi}$ means the noise predictor of the diffusion model, $\eta_{1}$ and $\eta_2$ are two predefined coefficients, and $y_{\text{pos}}$ and $y_{\text{neg}}$ are the positive and negative text prompts. In Fig.~\ref{fig:pipeline}, $y_{\text{pos}}$ represents the input text prompts, such as ``A turtle'', while $y_{\text{neg}}$ contains predefined negative text prompts such as ``oversaturated color'', ``ugly'', ``underexposed'', and ``overexposed''.}

However, simply applying the Stable Diffusion model here for distillation would lead to erroneous material parameters in two aspects. First, the generated materials may not be consistent with the given geometry. An example is shown in Fig.~\ref{fig:ablation} (a) because the diffusion model may generate images aligned with the prompt but not aligned with the geometry. Second, we render the image using a predefined environment light, but the generated images of the diffusion model may not be consistent with the given environment light as shown in Fig.~\ref{fig:light_aware}. Inconsistencies can result in materials with incorrect albedo, including baked-in shadows or highlights as shown in Fig.~\ref{fig:ablation} (b). To align images with text prompts, geometry, and environment light, we created a diffusion model that considers both geometry and light.

\subsection{Geometry- and Light-aware Diffusion Model}
\label{sec:diffusion}
We finetune the Stable Diffusion model by adding additional geometry and light conditions with ControlNet~\cite{zhang2023adding}. As shown in Fig.~\ref{fig:controlnet}, the geometry condition is the rendered depth and normal maps. To represent the light condition, we assign a set of predefined materials to the object, which have the same white albedo color but different metallic and roughness. Then, we render images of this object using the given environment light and these predefined materials, which are used as the light condition to the ControlNet. Finally, we finetune the ControlNet with both light and geometry conditions on the Objaverse~\cite{deitke2023objaverse} dataset. The output examples of the finetuned ControlNet are shown in Fig.~\ref{fig:controlnet}. The resulting diffusion model generates images that are consistent with the geometry and the lights, which are used to calculate CSD loss to generate high-quality materials.

\subsection{Material Generation}
\label{sec:generation}
In this section, we summarize the material generation process of DreamMat, as shown in Fig.~\ref{fig:pipeline}. 
Given the input mesh, we first precompute its light conditions on 128 random viewpoints using 5 predefined different environment lights because calculating the light conditions on the fly is time-consuming.
During each distillation step, we randomly select a viewpoint and an environment light to render an image on the mesh using the material representation as stated in Sec.~\ref{sec:inv}.
Then, we add noise to the rendered image. The noisy rendered image is used in the computation of the CSD loss stated in Sec.~\ref{sec:csd} with the geometry- and light-aware diffusion model in Sec.~\ref{sec:diffusion}, which uses the corresponding geometry and lighting condition of this viewpoint and this environment light. The CSD loss is backward to optimize the parameters in the material representation.
Except for the CSD loss, following the previous works~\cite{zhang2022invrender, yang2023sire, liu2023nero}, we apply a material smoothness loss
\begin{equation}
\mathcal L_{\text{smooth}}=||\Gamma_\theta(\mathbf{p})-\Gamma_\theta(\mathbf{p}+\mathbf{\epsilon})||^2,
\end{equation}
where $\mathbf{\epsilon}$ is a small random perturbation vector sampled from a Gaussian noise with 0.05 as its standard deviation. This smoothness loss $\mathcal L_{\text{smooth}}$ makes the predicted materials (roughness, metallic, and albedo) more smooth on the mesh. Finally, we sample the trained material representation $\Gamma_\theta$ to construct material UV maps for compatibility with modern rendering engines. 
\section{Experiments}
\begin{figure*}
  \includegraphics[width=0.99\linewidth]{{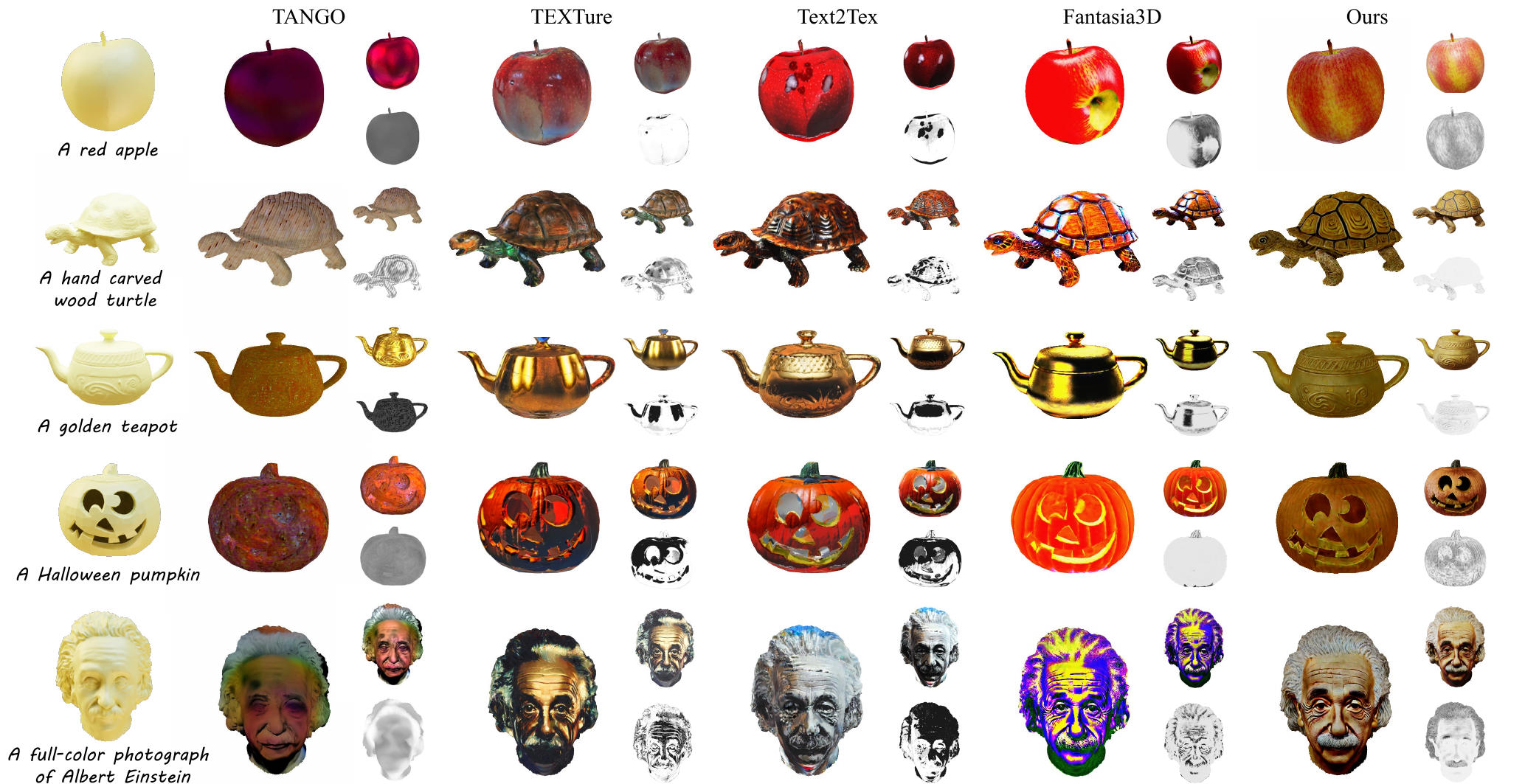}}
  \caption{\textbf{Qualitative comparison.} We compared our method to TANGO \cite{chen2022tango}, TEXTure \cite{yu2023texture}, Text2Tex \cite{chen2023text2tex}, and Fantasia3D \cite{Chen_2023_ICCV}. We use NvDiffRec \cite{Munkberg_2022_CVPR} to decompose the texture map produced by TEXTure and Text2Tex. Each object has three images: the albedo map on the left, the rendered image on the top right, and the roughness map on the bottom right.}
  \label{fig:comparison}
\end{figure*}

\begin{table}
\caption{\textbf{User study} conducted with 42 respondents. This table shows the average scores given by participants. We ask them to evaluate the albedo, roughness, metallic, and renderings of TANGO~\cite{chen2022tango}, TEXTure~\cite{yu2023texture}, Text2Tex~\cite{chen2023text2tex}, and Fantasia3D~\cite{Chen_2023_ICCV} to give scores in [1, 5], where a higher score means a better result. The evaluation criteria included overall quality, fidelity to the text prompt, effectiveness of albedo disentanglement from shading effects (``Light/Mat. Disen''), \revision{individual quality of the different material maps and quality of rendering under new environment lights rather than those used in generation}. Scores were averaged across all responses and examples. The questionnaire used in this user study is in the supplementary material.}
\vspace{-2mm}
\centering
\scalebox{0.99}{
\setlength{\tabcolsep}{2pt}
\begin{tabular}{lccccc}
\toprule

Method & \begin{tabular}[c]{@{}c@{}}TANGO \end{tabular} & \begin{tabular}[c]{@{}c@{}}TEXTure \end{tabular} & 
\begin{tabular}[c]{@{}c@{}}Text2Tex \end{tabular} & 
\begin{tabular}[c]{@{}c@{}}Fantasia3D \end{tabular} & 
 \begin{tabular}[c]{@{}c@{}}Ours\end{tabular} \\

\midrule
Overall Qual.     & 1.77 & 3.00 & 3.04 & 2.97  & \textbf{4.39}  \\
Text Fidelity       & 2.32 & 3.34 & 3.26 & 3.26  & \textbf{4.41}  \\
Albedo Qual.     & 1.67 & 2.92 & 2.73 & 2.95  & \textbf{4.65}  \\
Roughness Qual.   & 2.21 & 2.48 & 2.63 & 2.57  & \textbf{4.41}  \\
Metallic Qual.    & 1.75 & 2.71 & 2.75 & 3.01  & \textbf{4.53}  \\
Light/Mat. Disen.   & 1.95 & 2.52 & 2.49 & 2.92  & \textbf{4.36}  \\
Rendering Qual.   & 1.37 & 3.01 & 3.04 & 3.10  & \textbf{4.75}  \\
\bottomrule
\end{tabular}
}
\vspace{-5mm}
\label{tab:user_study}
\end{table}

\begin{figure*}
  \includegraphics[width=\linewidth]{{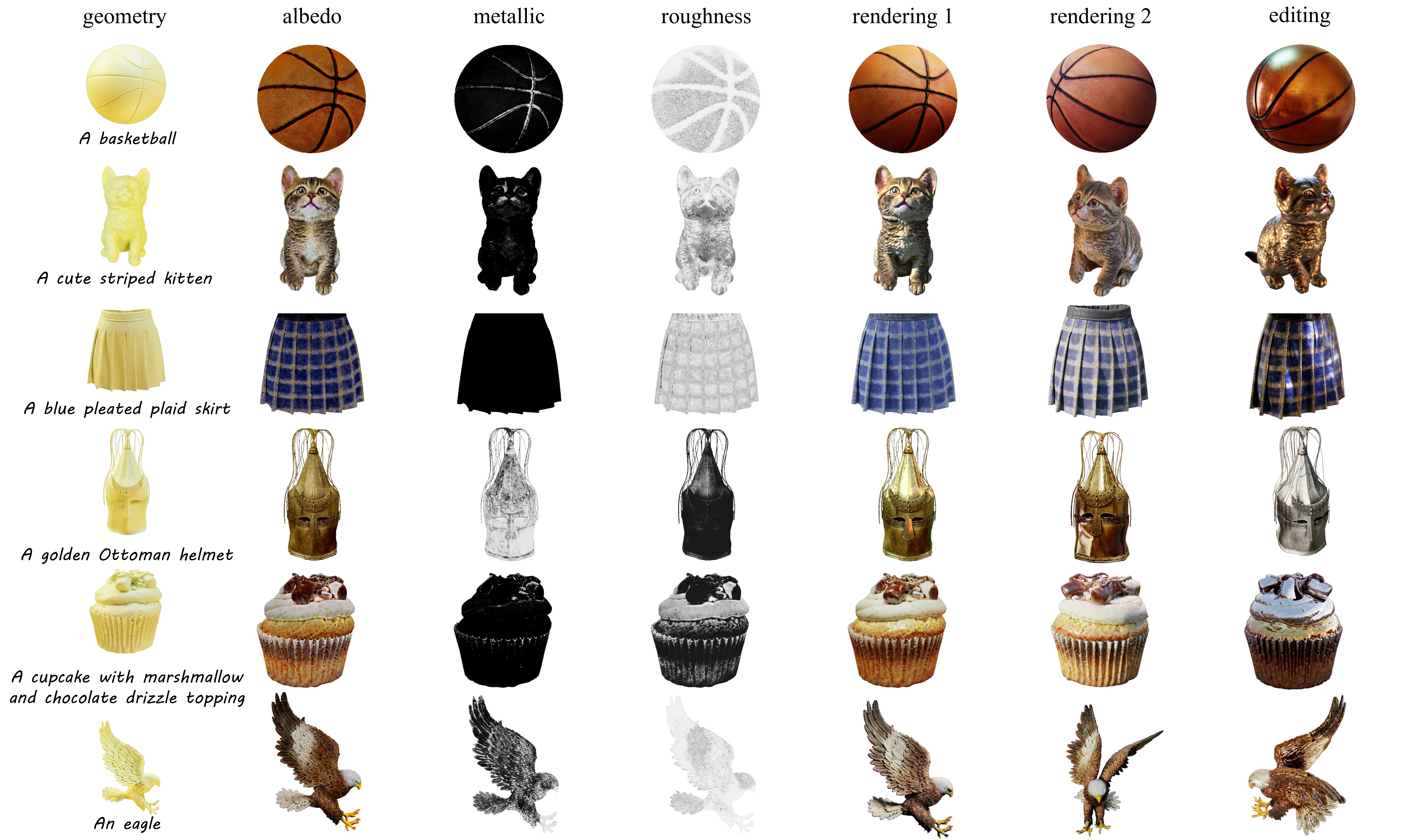}}
  \caption{\textbf{More generated materials and editing results}.}
  \label{fig:relight}
  \vspace{-3mm}
\end{figure*}

\subsection{Implementation Details}
\label{sec:implement}
We train the geometry- and light-aware ControlNet from the images which are rendered on the objects in the LVIS subset of the Objaverse \cite{deitke2023objaverse}. Since the names and tags of objects in this dataset are rather noisy, we employ BLIP \cite{li2022blip} for captioning all rendered images. Following \cite{liu2023syncdreamer}, we render 16 random views for every object under randomly chosen environment light maps. The light condition maps are obtained by using ray tracing in Blender, which represents the \revision{radiance} for different materials under the environment light. For normal maps, we transform the model's normal vectors into view space and flip the x-axis following ScanNet’s \cite{dai2017scannet} protocol. Depth maps are processed by inverting the real depth values and normalizing them. The filtered subset contains 1,242,880 entries, each consisting of conditional images and one rendered image. Our diffusion model is based on Stable Diffusion v2.1 and is trained with a batch size of 256 for 3 epochs, utilizing 8 V100 GPUs. More training details and results are presented in the appendix.

Our material generation pipeline is implemented in ThreeStudio~\cite{threestudio2023}. As rendering light conditions is time-consuming, we randomly sample 128 viewpoints and render images with five different environment lights using Blender. In each iteration, we randomly select a viewpoint and environment light, optimizing the materials for 4,000 steps using an Adam optimizer with learning rate 0.01. We set the control scale at 1.0 in the ControlNet with a gradual decay to 0.8 after 700 steps. Adhering to the CSD loss with annealed negative prompts, we set the $\eta_1=1.05$ and $\eta_2$ was progressively reduced from 1.0 to 0.5. 

\subsection{Qualitative Results}

\label{sec:comparison}
\subsubsection{Baselines}
We compare our method with several state-of-the-art methods for 3D appearance generation, namely TANGO \cite{chen2022tango}, TEXTure~\cite{richardson2023texture}, Text2Tex \cite{chen2023text2tex}, and Fantasia3D~\cite{Chen_2023_ICCV}. Since TEXTure and Text2Tex can only generate a single RGB texture map, we utilize NvDiffRec~\cite{Munkberg_2022_CVPR} to post-process the generated textures for material decomposition. As the mesh is given, we employ only the second stage of Fantasia3D to optimize the materials with the fixed geometry, normal, and environment light map. 
\subsubsection{Comparison with baselines}
We take diverse 3D meshes from real game assets and the Objaverse~\cite{deitke2023objaverse} dataset that are not included in the training set. Fig.~\ref{fig:comparison} shows the albedo and roughness generated from the same text prompts and untextured meshes. We also show the renderings of the generated materials under the same environment light. Tango successfully recognizes specific material descriptors like ``golden'' and ``wooden'' but falls short in generating detailed textures for objects. TEXTure and Text2Tex only generate pure RGB textures instead of PBR materials. Although the inverse rendering technique is employed for decomposing texture maps into materials, we are still unable to correctly disentangle albedo from environmental lights. Fantasia3D is capable of directly generating PBR materials but tends to retain many highlights and shadows in the albedo, due to the absence of lighting constraints in its generation process. Our method stands out by effectively achieving disentanglement of materials from environmental lighting with material diversity and texture detail preserved.

\subsubsection{User study}
To further validate the stability and quality of our model, we conduct a user study on the 20 generated materials. Each participant is provided with 5 examples, accompanied by the corresponding text prompts and meshes. They rated the materials generated by each method on various aspects. 42 feedbacks from 37 users were collected from the study with results detailed in Tab.~\ref{tab:user_study}.

\subsubsection{More results and editing}
The high-quality PBR material decoupled from lighting enables photo-realistic rendering in a modern graphics engine like Blender. Fig.~\ref{fig:relight} shows more results with the albedo, roughness, and metallic material generated for a given 3D model from text prompts, as well as their renderings under different environment lights. 
Moreover, by adjusting the overall roughness or metallic, or by altering the overall hue of the albedo, we can conveniently edit the material to achieve different visual appearances.

\subsection{\revision{Quantitative Comparison}}
\revision{To evaluate the usability and robustness of DreamMat, we conduct a quantitative analysis in comparison with baseline methodologies using CLIP score~\cite{hessel-etal-2021-clipscore} and FID~\cite{heusel2017gans} of rendered images as metrics. Our experiment involves generating materials for 10 different 3D meshes and selecting 5 distinct text prompts for each mesh for generation. To compute the metrics, we randomly chose 120 viewpoints for each object to render images. Then, the semantic alignment between the text prompts and the rendered images is quantitatively assessed by the CLIP Score~\cite{hessel-etal-2021-clipscore}, where a higher score indicates a greater similarity between the generated appearance and the text prompts. Furthermore, the quality of the rendered images is evaluated by the Fréchet Inception Distance (FID)~\cite{heusel2017gans}, which compares the distribution’s distance between the images generated by Stable Diffusion and the rendered images from the generated appearances. As illustrated in Tab.~\ref{tab:quantitative}, DreamMat outperforms the baseline methods in achieving the best text fidelity and superior visual quality of the generated appearances.}

\subsection{Ablative Study}
\label{sec:ablation}

To validate the effectiveness of each component, we conduct an ablation study on the text prompt ``a wooden treasure chest with metal accents and locks'' to generate materials for a given untextured treasure chest mesh. The results of the ablation study are shown in Fig.~\ref{fig:ablation} and more ablation study results on other meshes and prompts are included in the supplementary material. 

\begin{table}
\caption{\revision{\textbf{Quantitative results.} We use 50 different text prompts on 10 meshes to generate appearances and calculate the CLIP score (similarity between rendered views and text prompts) and FID (distribution’s distance between rendered images from the generated appearances and the generated images by Stable Diffusion) to assess the text fidelity and visual quality of the generated appearances.}}
\centering
\scalebox{0.95}{
\setlength{\tabcolsep}{3pt}
\begin{tabular}{lccccc}
\toprule

Method & \begin{tabular}[c]{@{}c@{}}TANGO \end{tabular} & \begin{tabular}[c]{@{}c@{}}TEXTure \end{tabular} & 
\begin{tabular}[c]{@{}c@{}}Text2Tex \end{tabular} & 
\begin{tabular}[c]{@{}c@{}}Fantasia3D \end{tabular} & 
 \begin{tabular}[c]{@{}c@{}}Ours\end{tabular} \\

\midrule
CLIP Score $\uparrow$    & 76.15 & 78.55 & 78.52 & 77.30  & \textbf{80.28}  \\
FID  $\downarrow$     & 165.40 & 135.72 & 144.86 & 131.86  & \textbf{114.97}  \\
\bottomrule
\end{tabular}
}

\label{tab:quantitative}
\vspace{-15pt}
\end{table}

\begin{enumerate}[noitemsep,nolistsep,leftmargin=*]
    \item \textbf{Baseline distillation method}. In Fig.~\ref{fig:ablation} (a), we directly combine our inverse rendering method with a text-to-image Stable Diffusion model to generate materials. Though some reasonable results are achieved in this baseline method, there is a noticeable inconsistency between the generated material and the given geometry as highlighted by the handle region of the chest.
    \item \textbf{Distillation with geometry-aware diffusion models}. An effective way to ensure the consistency between materials and geometry is to adopt the depth and normal-conditioned diffusion model. Fig.~\ref{fig:ablation} (b) shows the distilled materials from a pre-trained normal and depth ControlNet~\cite{zhang2023adding}. This distillation method yields more consistency between the generated materials and the input mesh's geometry, as evidenced by the more pronounced texture details. 
    However, because of the ill-posed nature of material decomposition, the generated albedo contains shading effects like highlights and the roughness and metallic are of low quality.
    \item \revision{\textbf{Distillation with light-aware diffusion models}. Fig.~\ref{fig:ablation} (c) shows the distilled materials from a light-aware ControlNet without the use of geometry conditions, which results in geometry inconsistency. Although lighting conditions already carry geometric clues, primarily through shadows, when there are not adequate shadows, only using light conditions is not enough to capture accurate geometry.}
    \begin{figure}
  \includegraphics[width=0.99\linewidth]{{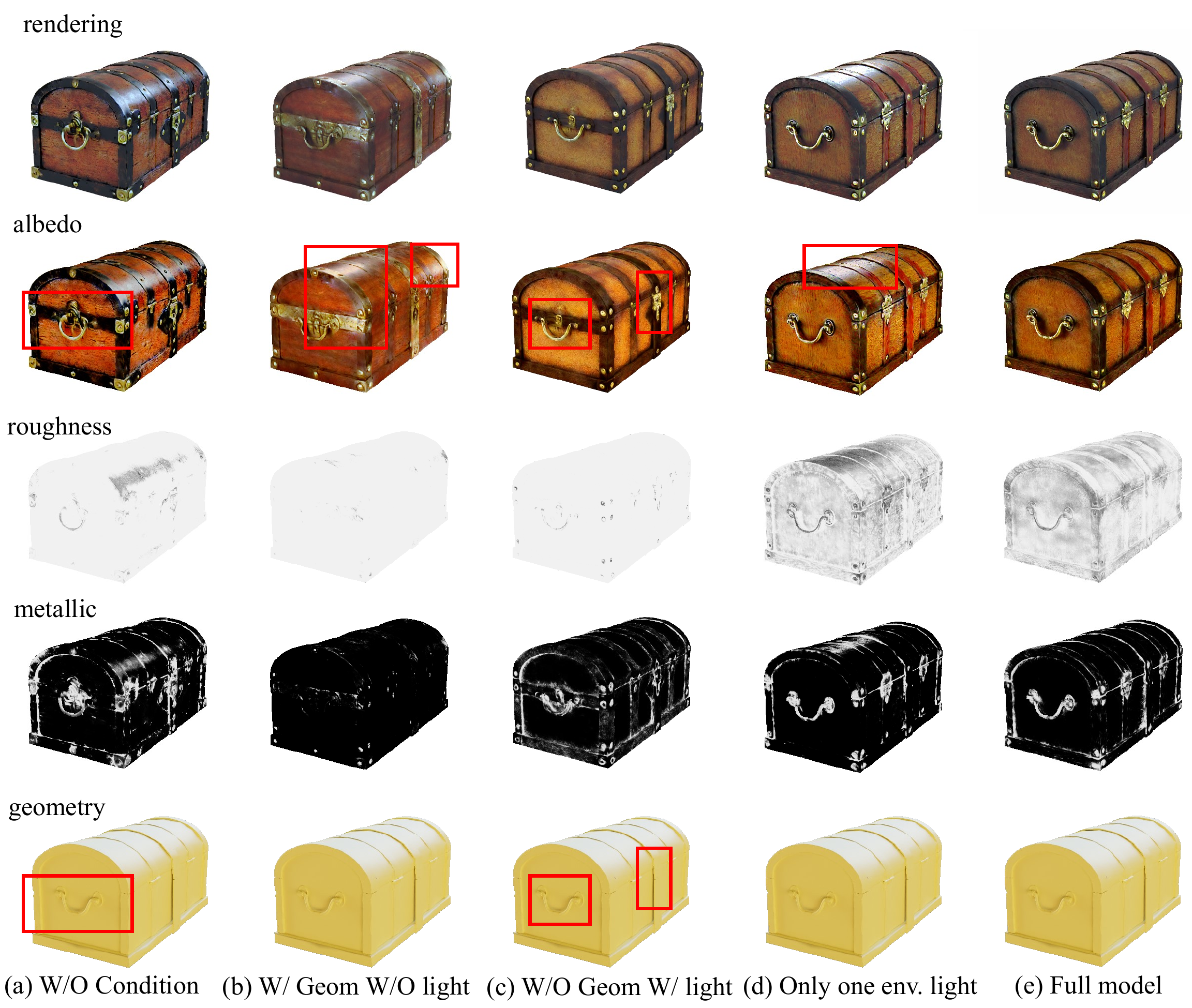}}
  \caption{\textbf{Ablation study} on the text prompt ``a wooden treasure chest with metal accents and locks'' applied to an untextured treasure chest mesh. (a) Baseline distillation method on a Stable Diffusion method with our inverse rendering scheme. (b) Adding ``Geom'' conditions, i.e. normal maps and depth maps, enables geometry-consistent material generation. \revision{(c) Employing solely lighting conditions in the absence of geometric constraints.} (d) Distilling our geometry- and light-aware diffusion model but only using one environment light in the distillation. (e) Our full model with geometry- and light-aware diffusion model and randomly selected environment light.}
  \label{fig:ablation}
  \vspace{-2mm}
\end{figure}
    \item \textbf{Fixed environment light with geometry- and light-aware diffusion model}. In Fig.~\ref{fig:ablation} (d), we apply the proposed geometry- and light-aware diffusion model to distill the material. However, instead of randomly selecting an environment light during the distillation, we always choose the same environment light. The results show that applying the geometry- and light-aware diffusion model improves the quality of the generated materials with better roughness and metallic while using the same environment light leads to overfitting in the inverse rendering. Thus, the resulting albedo still contains incorrect highlights.
    \item \textbf{Full model}. Fig.~\ref{fig:ablation} (e) shows the results of our full method, which distills the geometry- and light-aware diffusion model and randomly selects an environment light in the distillation. Introducing the light condition enables the generation of light-consistent images. Meanwhile, by varying the light maps across iterations, we prevent the material's properties from overfitting to one environment light and remove the shading effects in the albedo. This produces the materials of the best quality and photorealistic shading results.
\end{enumerate}

\begin{figure*}
  \includegraphics[width=0.95\linewidth]{{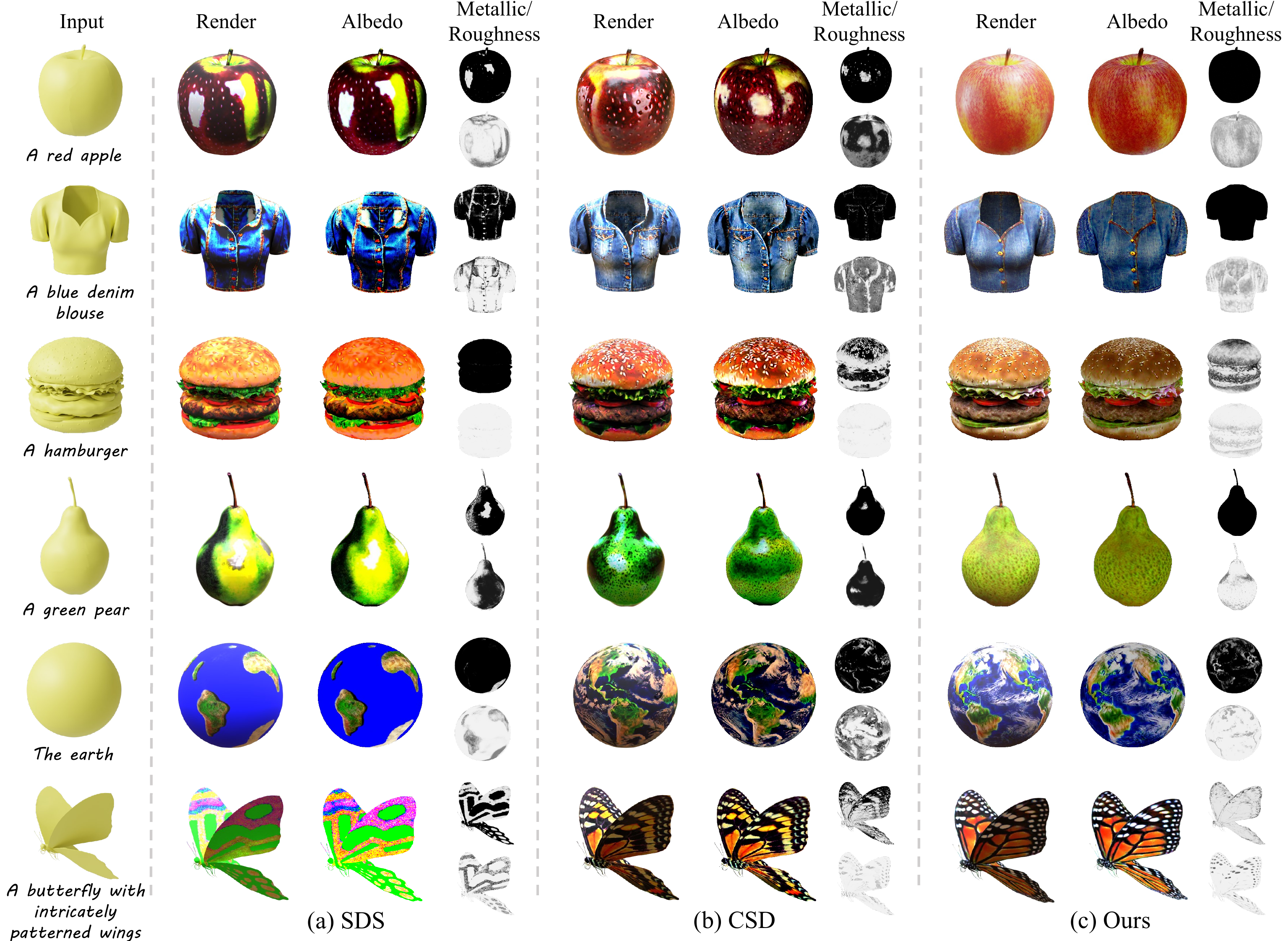}}
  \caption{\revision{\textbf{Comparison between the vanilla SDS loss, the vanilla CSD loss, and our method.} (a) Materials generated with vanilla SDS loss~\cite{poole2022dreamfusion} and geometry condition. (b) Material generated using vanilla CSD loss~\cite{yu2023csd} with geometry condition. (c) Materials generated by our method which combines the CSD loss with our geometry- and light-aware diffusion model.}}
  \label{fig:sds_csd}
\end{figure*}
\subsection{Discussion on distillation losses}
In Fig.~\ref{fig:sds_csd}, we show the effects of using different distillation losses and different diffusion models. In Fig.~\ref{fig:sds_csd}(a), we adopt the SDS loss as DreamFusion~\cite{poole2022dreamfusion}, incorporating a ControlNet with Canny edges and depth for distillation. In Fig.~\ref{fig:sds_csd}(b), the CSD loss, following the texture generation approach described in ~\cite{yu2023text}, also employs the ControlNet with depth and Canny edges as conditions. The material representation and rendering approach remains the same as DreamMat. CSD loss significantly reduces oversaturation and shows more details compared to SDS loss, yielding a more realistic appearance. However, CSD alone tends to incorporate lighting information into the albedo. Our solution, as shown in Fig.~\ref{fig:sds_csd} (c), introduces a geometry- and light-aware diffusion model that better separates material properties from lighting, enhancing geometric fidelity. We also include results of using SDS loss with our geometry- and light-aware model in the Appendix~\ref{sec: sds_appendix}.

\subsection{Generating Diverse Materials}
\begin{figure*}
  \includegraphics[width=\linewidth]{{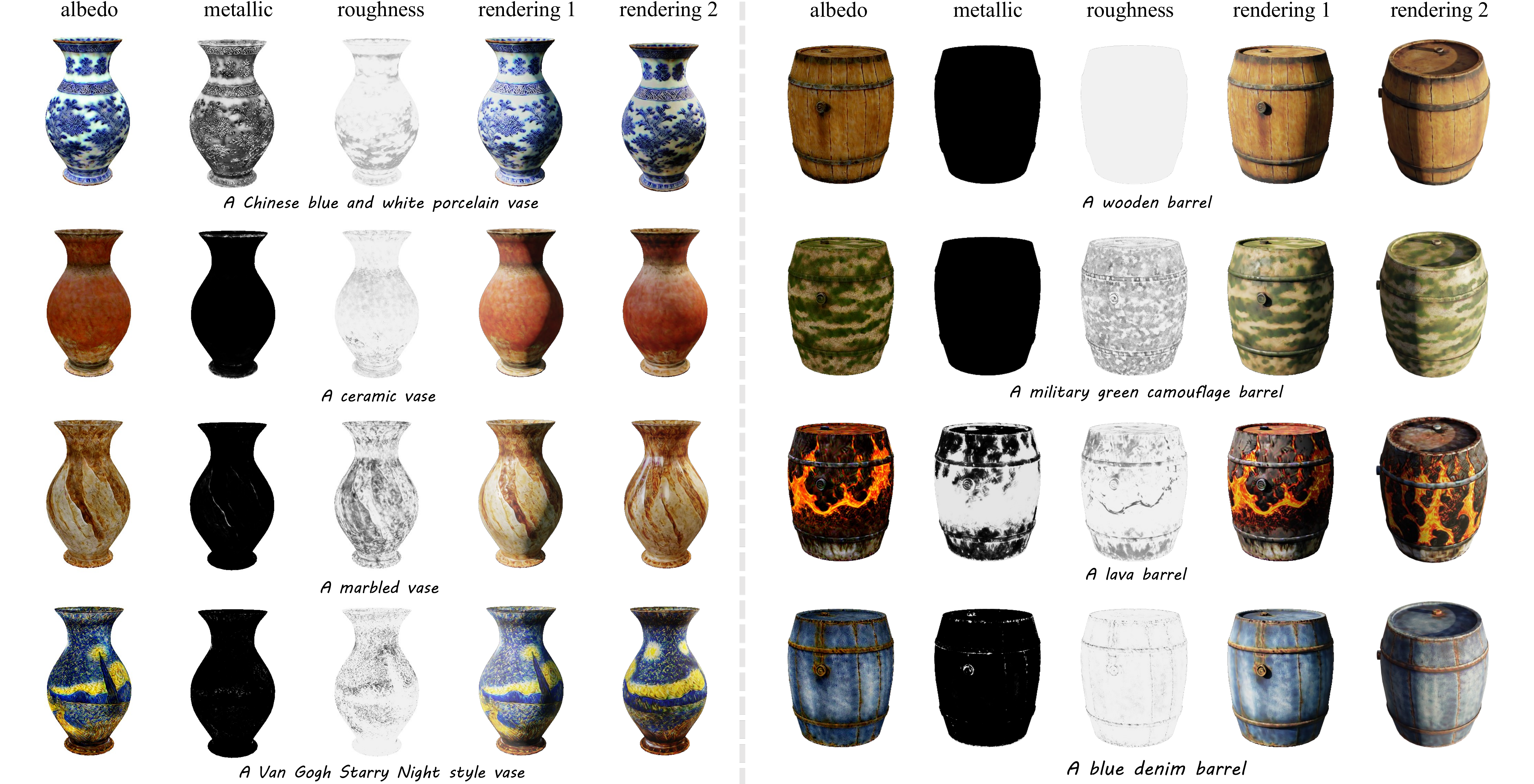}}
  \caption{\textbf{Diverse material generation}. Our method can generate different materials with different text prompts on the same mesh.}
  \label{fig:diversity}
  \vspace{-1mm}
\end{figure*}
Our method is able to generate different materials given different prompts on the same mesh. Fig.~\ref{fig:diversity} shows the materials and the rendering results of the same mesh generated from different text prompts by the proposed methods. 
These results demonstrate that our method is able to generate diverse materials of different styles that align well with the text prompts with high fidelity.
\subsection{Material Generation of Complex Objects}
\begin{figure*}
  \includegraphics[width=\linewidth]{{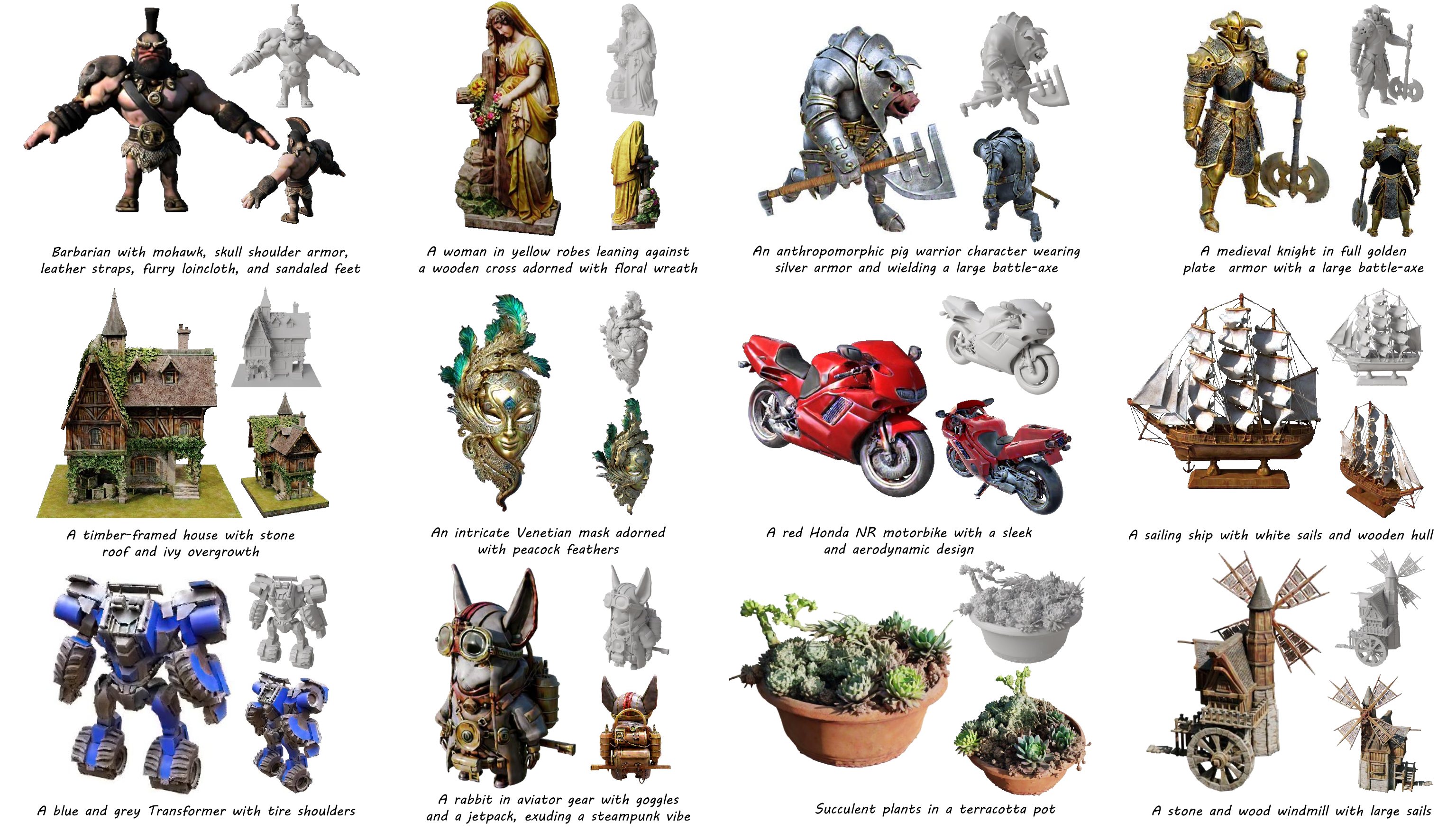}}
  \caption{\textbf{Material generation of complex objects}. On the given complex meshes shown in the right-top, our method can generate high-quality materials for the whole mesh in one distillation process, which enables photo-realistic renderings in Blender (left and right-bottom parts).}
  \label{fig:complex}
\end{figure*}

We demonstrate the generative capabilities of our method on a set of challenging examples, as shown in Fig.~\ref{fig:complex}, which include complex self-occlusions, assemblies of multiple parts, and a variety of material compositions. Our method can directly take the whole mesh as input and generate textures on the mesh according to the given text prompt. Then, we export the generated materials into 2048x2048 resolution albedo, roughness, and metallic texture maps, which are utilized in the Blender to render the photo-realistic images in Fig.~\ref{fig:complex}.

\begin{figure*}
  \includegraphics[width=\linewidth]{{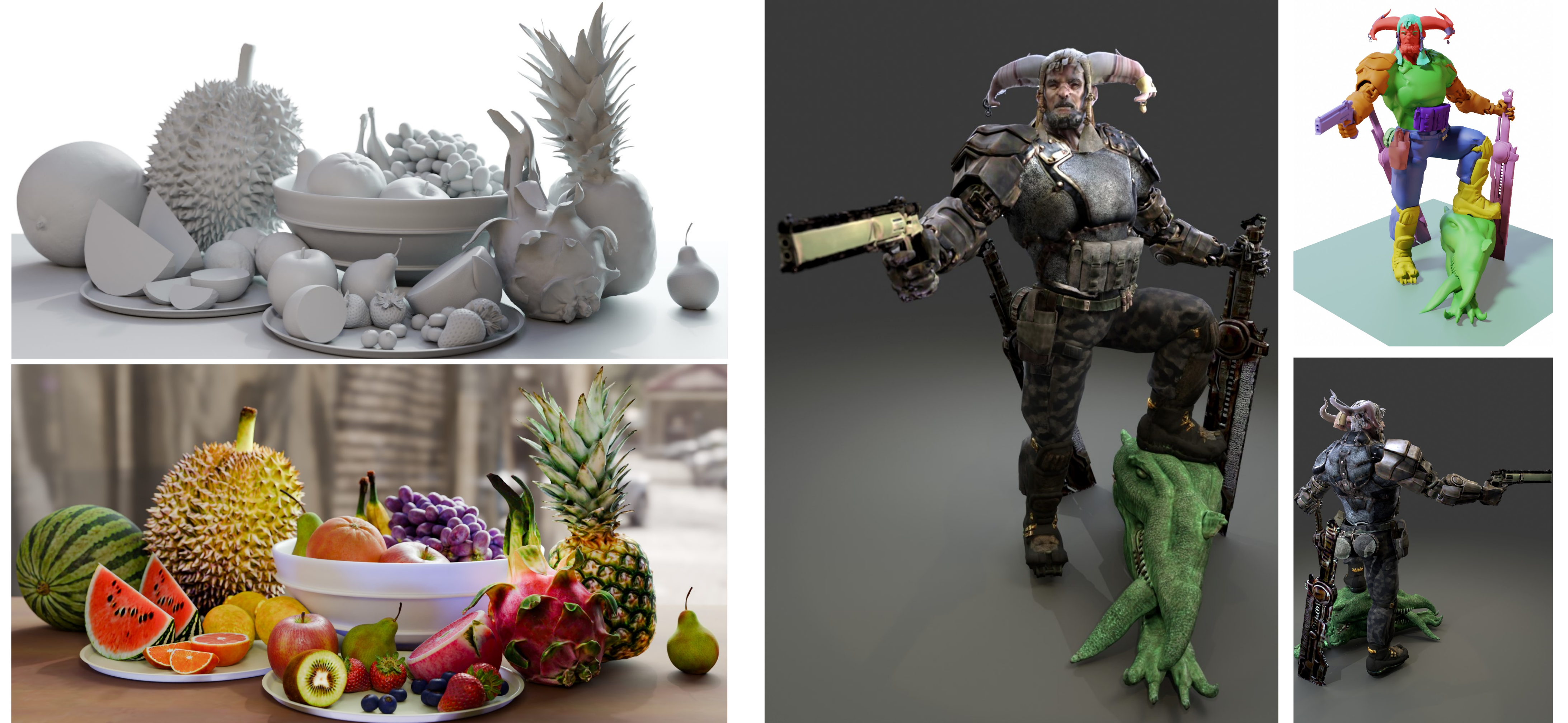}}
  \caption{\textbf{Combining generated materials of different components}. For extremely complex examples, our method can generate materials for each component of the meshes separately and then combine them to get the materials for the whole mesh.
  }  
  \label{fig:fruit}
  \vspace{-3mm}
\end{figure*}
\subsection{Material Generation for Object Sets}
For a cluttered scene or a highly-detailed avatar, as shown in Fig.~\ref{fig:fruit}, it is difficult to directly generate materials for the whole scene or the whole body in one distillation process. Instead, we generate materials for each component separately and then combine these parts together to get the generated appearances for these two extremely complicated examples. For the ``fruit'' scene, we generate materials for each fruit separately and combine them. For the ``avatar'' mesh, the top-right figure shows the separated components with different colors. Utilizing ray tracing for rendering, our demonstration showcases the compatibility of our method with modern computer graphics rendering pipelines. The results achieved are of photo-realistic quality, illustrating the effectiveness of our approach in producing visuals that closely mimic real-world appearances. 

\subsection{Runtime Analysis}
We conduct a runtime analysis of our method on one NVIDIA RTX 4090 graphics card. For the ``Transformers'' mesh displayed in Fig.~\ref{fig:complex}, our material distillation process costs about 18 minutes. The rendering and optimization of the material cost about 2/3 time (12 min) while querying the diffusion model costs about 1/3 time (6 min).

\section{Limitations and Conclusions}
\subsection{Limitations}
Although our method successfully generates diverse and high-quality appearances according to the text prompts, it still has several limitations. \revision{Due to the ill-posed nature of material decomposition, DreamMat still exhibits inaccurate metallic and roughness in some cases. Also, our method has difficulty} in dealing with materials that exhibit properties like transparency, high reflection, or subsurface scattering.
This limitation is primarily caused by our choice of the Bidirectional Reflectance Distribution Function (BRDF) model, which cannot model more advanced and complex material. Meanwhile, the proposed method only accounts for the direct lights from the environment map but does not consider the indirect lights reflected from the object itself, which may lead to incorrect materials for highly reflective objects. Considering the indirect lights and more advanced BRDF may resolve this limitation but also bring more computation complexity in distillation, which we leave for future works. Another limitation is that the distillation of our method takes a relatively long time for a high-quality generation (about 20 minutes) while designers may want to use the material generation in an interactive environment. Our method may be further sped up with recent faster diffusion models and advanced representations in future works. \revision{Since DreamMat is based on Stable Diffusion~\cite{rombach2022high}, it shares some limitations of Stable Diffusion, such as difficulty in precisely controlling the individual material components solely through text prompts.}
\subsection{Conclusions}
In summary, we present a novel text-guided technique for generating detailed PBR materials specifically for given untextured 3D meshes. Our method includes a geometry- and light-aware diffusion model and an inverse rendering-based distillation method. The inverse rendering method renders images by Monte Carlo sampling and distills materials by a CSD loss. The key advantage of our method is the geometry- and light-aware diffusion model which can generate images consistent with the geometry and environment light. Distilling from this diffusion model avoids the common problem of baking shading effects into albedo. We demonstrate that the generated materials by our method are readily usable in modern graphics engines, offering enhanced realism for various applications in gaming and simulation.
\begin{acks}
Xiaogang Jin was supported by the Key R\&D Program of Zhejiang (No. 2023C01047) and the FDCT under Grant 0002/2023/AKP. 
This research work was supported by Information Technology Center and State Key Lab of CAD\&CG, ZheJiang University.
\end{acks}

\bibliographystyle{ACM-Reference-Format}
\bibliography{main}

\appendix
\section{Appendix}
\begin{figure*}
\includegraphics[width=\linewidth]{{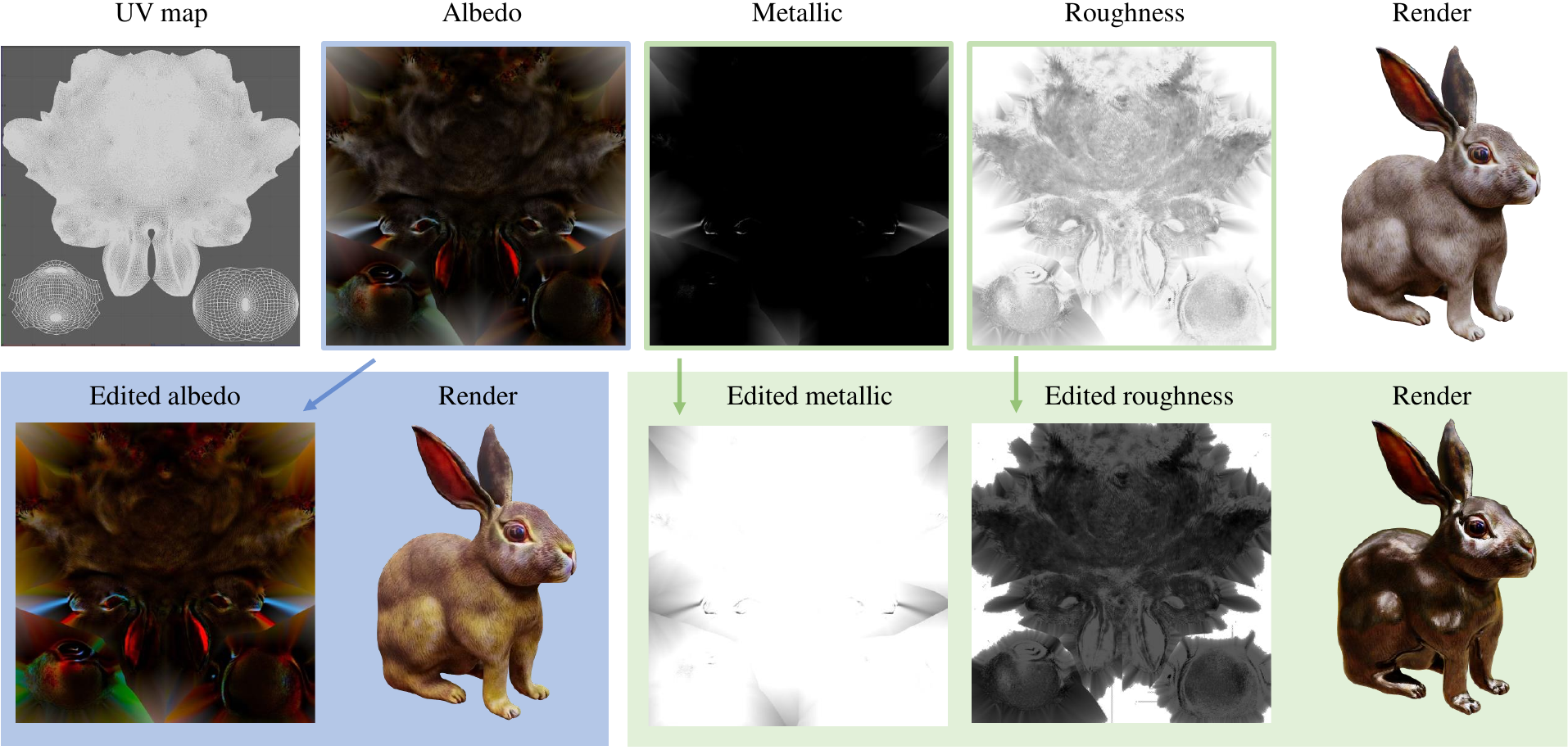}}
  \caption{\revision{\textbf{Texture maps and material editing.} Top: Generated material maps utilizing DreamMat and the rendering results. Bottom: Edited material maps using 2D image editing techniques and their rendering results.}}
  \label{fig:uv}
\end{figure*}

\subsection{More Implementation Details}

\begin{figure*}
  \includegraphics[width=\linewidth]{{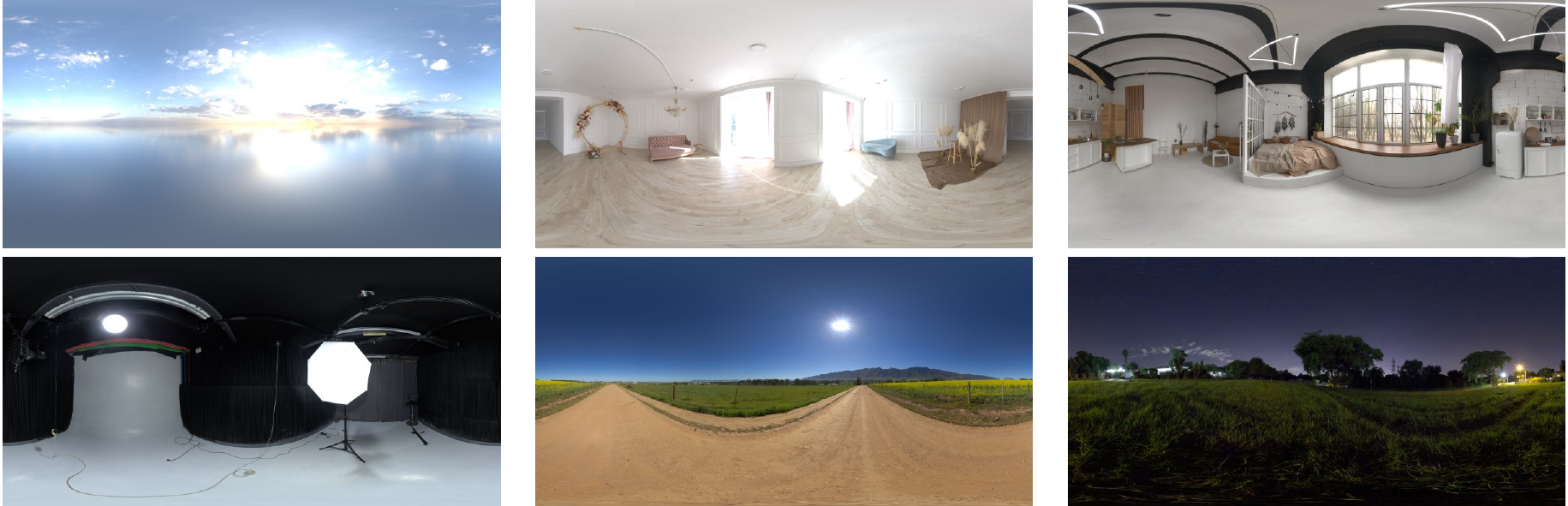}}
  \caption{Environment light maps used in geometry- and light-aware diffusion model training and material distillation.}
  \label{fig:environment}
\end{figure*}

\textbf{Geometry- and Light-aware diffusion model Training.}
The geometry- and light-aware diffusion model is trained using six distinct environment light maps, as illustrated in Fig. \ref{fig:environment}. Within the Objaverse dataset \cite{deitke2023objaverse}, we exclude objects that failed to meet specific criteria, such as those containing transparent parts, non-mesh structures, or too simple geometry (such as only a single plane). Then, we randomly choose one environment light map to render both the light condition maps and the final results. During the rendering of the light condition map, we set albedo to white color, metallic to 0.0 and 1.0, and roughness to 0.0, 0.5, and 1.0. 

\begin{figure*}
  \includegraphics[width=\linewidth]{{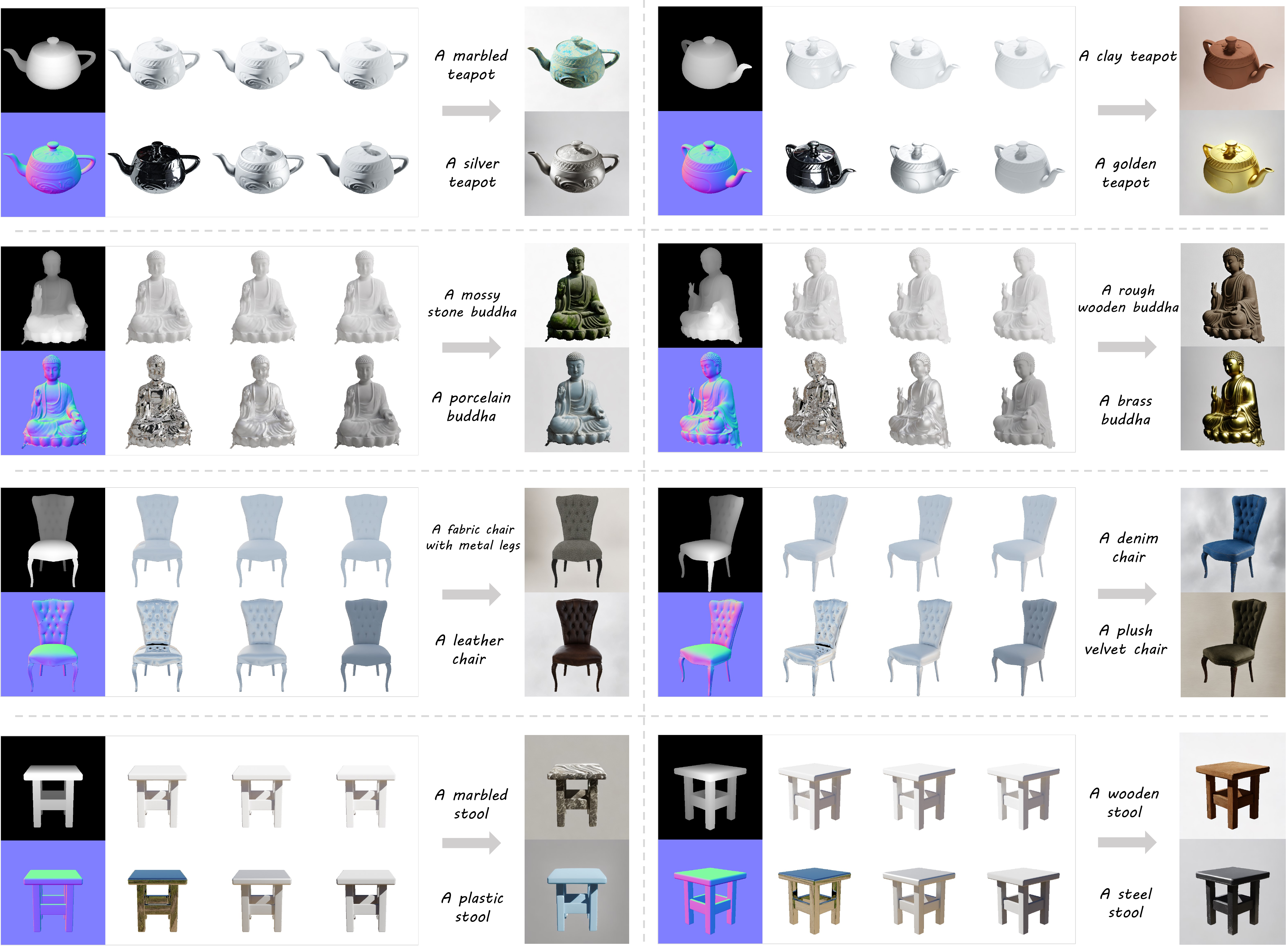}}
  \caption{Image generated by the geometry- and light-aware diffusion model with different text prompts.}
  \label{fig:controlnet_res}
\end{figure*}

Our diffusion model uses a ControlNet~\cite{zhang2023adding} with a condition map of 22 channels. The learning rate is set as 1e-5 with a single-step gradient accumulation and a batch size of 256. The model is trained for a total of 3 epochs, which uses 8 V100 GPUs for 3 days. To demonstrate the versatility and quality of the ControlNet, Fig. \ref{fig:controlnet_res} illustrates the generated results using different text prompts under different lighting conditions.

\textbf{UV Mapping and Material Editing}
During the texture map output phase, we employ UV mapping to sample the generated appearance. The model's inherent UV map can be utilized, or alternatively, a UV map can be automatically generated using xatlas~\cite{xatlas}. Following the previous methodologies~\cite{Munkberg_2022_CVPR, Chen_2023_ICCV}, we apply the UV edge padding technique to extend the boundaries of UV islands and fill in empty regions. The output texture maps are shown in Fig.~\ref{fig:uv}, which can be seamlessly integrated into graphics engines.

Furthermore, these texture maps can be imported into various image editing software (e.g., Photoshop) for material editing. As demonstrated in the bottom row of Fig.~\ref{fig:uv}, we adjust the albedo's saturation, invert the metallic properties, and modify the overall brightness of the roughness. These adjustments enable the achievement of diverse rendering outcomes under the same lighting conditions.

\textbf{Baseline implementation details.}
For TANGO~\cite{chen2022tango}, we follow the official implementation and set the learning rate to 5e-4, which is decayed by 0.7 in every 500 iterations. We iterate 1500-3000 times for each object until convergence. Since TANGO uses position encoding $\beta(l) =  [\cos(2\pi Bl), \sin(2\pi Bl) ]^T$ to provide high-frequency details of the generated materials, where B is a random Gaussian matrix whose entry is randomly drawn from $N (0, \sigma^2)$ (different examples in the open source code use different position encoding parameters), we find that the final result is greatly influenced by the PE parameters. Therefore, we carefully tune the $\sigma$ and frequency number for each example to achieve the best performance. 

For TEXTure~\cite{richardson2023texture} and Text2Tex~\cite{chen2023text2tex}, we use their original implementations to generate texture results, and then apply NVdiffrec~\cite{Munkberg_2022_CVPR} for subsequent material decomposition. In the material decomposition stage, we fix the geometry and lighting with a learning rate of 0.01 for 3000 iterations.

For Fantasia3D~\cite{Chen_2023_ICCV}, we only use its second stage to generate the objects’ materials. In this stage, we fix the environmental lighting and geometry, optimizing the albedo, roughness, and metallic with prompts related to the viewpoint for 3000 iterations. The learning rate is set to 0.01.

\subsection{Condition Maps to CSD Loss}

\begin{figure}
  \includegraphics[width=1\linewidth]{{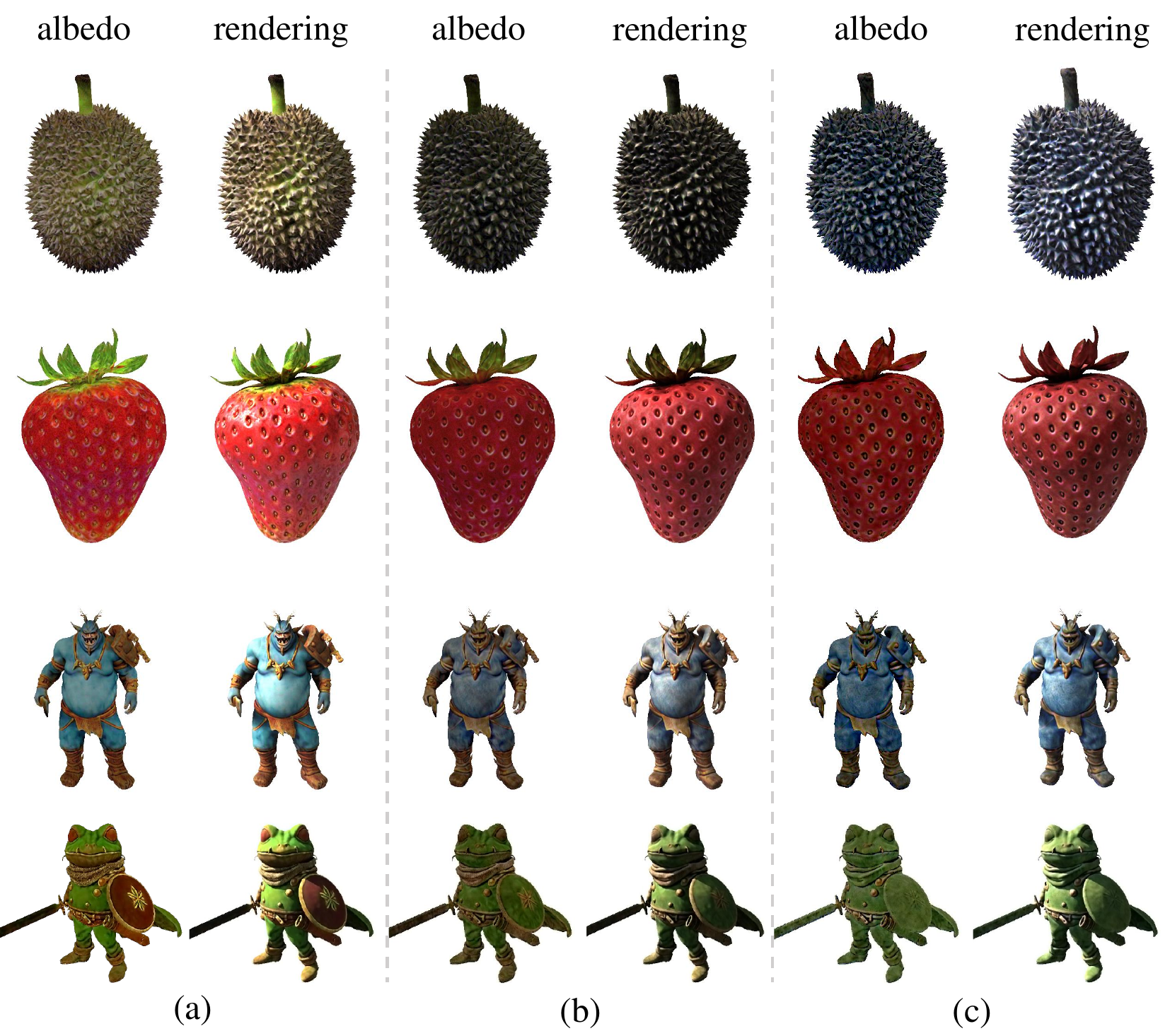}}
  \caption{(a) Our result, incorporating condition maps concurrently to $\epsilon_{\phi}(I_t; y_{\text{pos}}, t)$ ,  $ \epsilon_{\phi}(I_t; t) $, $\epsilon_{\phi}(I_t; y_{\text{neg}}, t)$; (b) The result when condition maps are only added to $ \epsilon_{\phi}(I_t; y_{\text{pos}}, t)$, $\epsilon_{\phi}(I_t; y_{\text{neg}}, t)$; (c) The result with condition maps only added into $\epsilon_{\phi}(I_t; y_{\text{pos}}, t)$}
  \label{fig:csd_supp}
\end{figure}

A proper control strength would be important in the CSD distillation process. As written in Sec 3.2, $\delta(I_t)$ consists of three components: $ \epsilon_{\phi}(I_t; y_{\text{pos}}, t)$, $ \epsilon_{\phi}(I_t; t) $, and $\epsilon_{\phi}(I_t; y_{\text{neg}}, t)$, representing the scenarios with positive prompts, without prompts, and with negative prompts, respectively. Following the implementation in the CSD~\cite{yu2023csd} source code, we applied the condition map to all three components, resulting in the outcomes shown in Fig.\ref{fig:csd_supp} (a). Additionally, we experimented with applying the condition map only to $ \epsilon_{\phi}(I_t; y_{\text{pos}}, t)$ and $\epsilon_{\phi}(I_t; y_{\text{neg}}, t)$ and found that similar results could be achieved, but with some color distortion, as shown in Fig. \ref{fig:csd_supp}(b). Moreover, applying the condition map solely to $ \epsilon_{\phi}(I_t; y_{\text{pos}}, t)$ suffers from a similar problem of tonal distortion.

\subsection{Entanglement of Materials and Lighting}
\begin{figure}
  \includegraphics[width=\linewidth]{{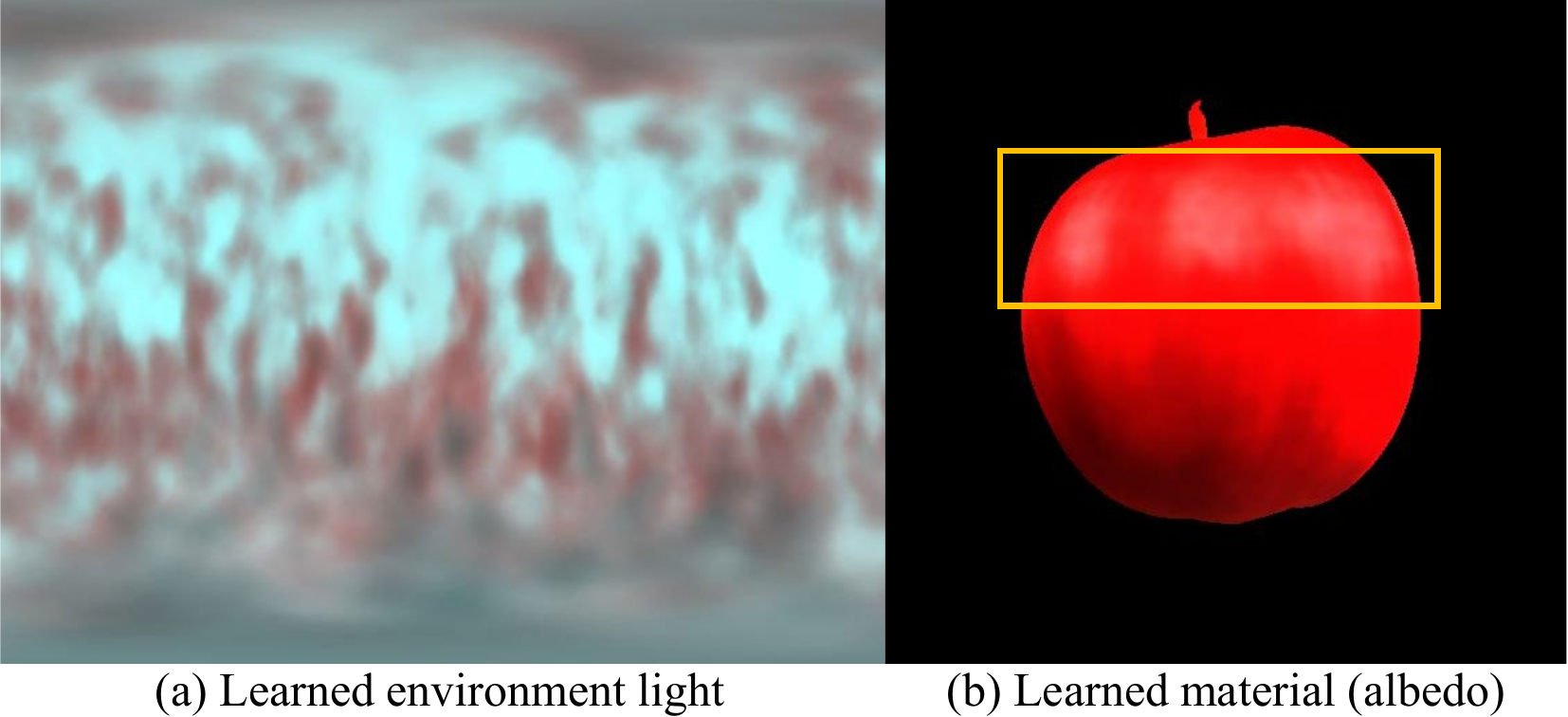}}
  \caption{\textbf{Results by directly distilling both materials nad environment light with an SDS~\cite{poole2022dreamfusion} loss on a stable diffusion~\cite{rombach2022high} model.} (a) the environment lighting map derived from the SDS optimization process, where material attributes of the apple are inadvertently encoded. (b) the albedo of the apple, which exhibits shading effects including highlights and shadows, indicative of the entanglement with the environmental lighting. This visual evidence supports the need for employing light maps to mitigate the ill-posed challenges of inverse rendering.}
  \label{fig:entanglement}
\end{figure}
In Fig.~\ref{fig:entanglement}, we show the results of directly optimizing both lighting and materials using SDS losses, without imposing any constraints on the diffusion process or the materials themselves. For lighting representation, we adopt the model used in NeRO~\cite{liu2023nero}, outputting the results as an environment lighting map after sampling. Notably, we observe that the learned environmental lighting contains substantial information about the material properties of the apple, while the apple's albedo retains numerous shading effects, such as highlights and shadows. This phenomenon stems from the ill-posed nature of inverse rendering, a challenge that becomes more pronounced when the lighting in stable diffusion-generated images is arbitrary. Consequently, the utilization of known environmental light maps is crucial to address these issues.
\begin{figure}
  \includegraphics[width=\linewidth]{{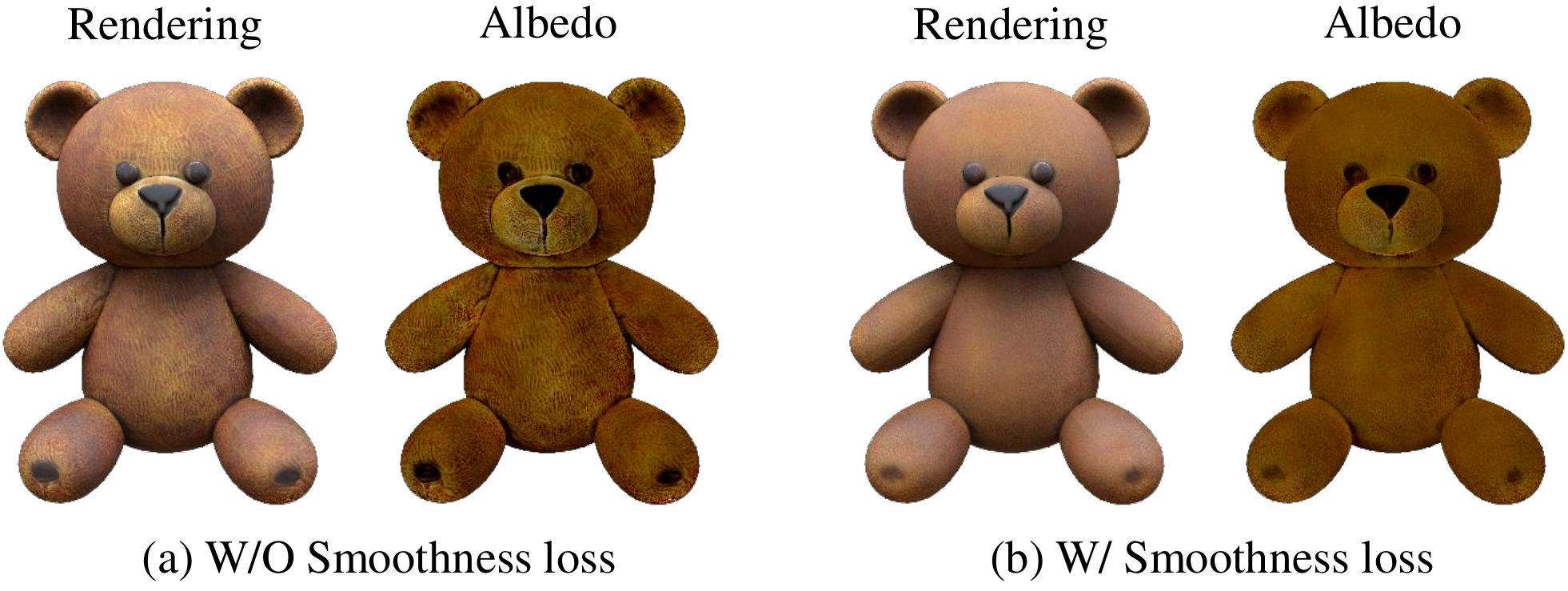}}
  \caption{\revision{\textbf{Ablation study} on the smoothness loss.}}
  \vspace{-2mm}
  \label{fig:smooth}
\end{figure}
\subsection{Different shading models in Objaverse}
\revision{In our material generation process, we employ the Disney BRDF. However, the Objaverse~\cite{deitke2023objaverse} dataset contains a variety of shaders within its .glb files, which results in a domain gap between renderings of some objects and their corresponding light condition maps during the training of our geometry- and light-aware diffusion model. However, we observe that this domain gap does not severely affect the quality of the generated materials of DreamMat. The reason is that we have excluded objects within Objaverse tagged with descriptors such as `stylized,' `style,' `handpainted,' `pixelart,' `npr,' and `non-photorealistic', which shows a large deviation from our simplified Disney BRDF. Some remaining objects still do not have the same shading models as ours. However, the effect of lighting on rendering showed consistent patterns as our shader. For example, highlights in renderings are produced by intense lighting. Thus, this still enables us to train our geometry- and light-aware diffusion model.
}

\subsection{Effects of smoothness loss}
\revision{
Following the previous methodologies, we incorporate a material smoothing loss, with the results depicted in Fig.~\ref{fig:smooth}. The inclusion of this smoothing term has resulted in a more refined albedo for the teddy bear. Users can adjust this term's coefficient to fine-tune the texture detail's granularity.}

\begin{figure}
  \includegraphics[width=\linewidth]{{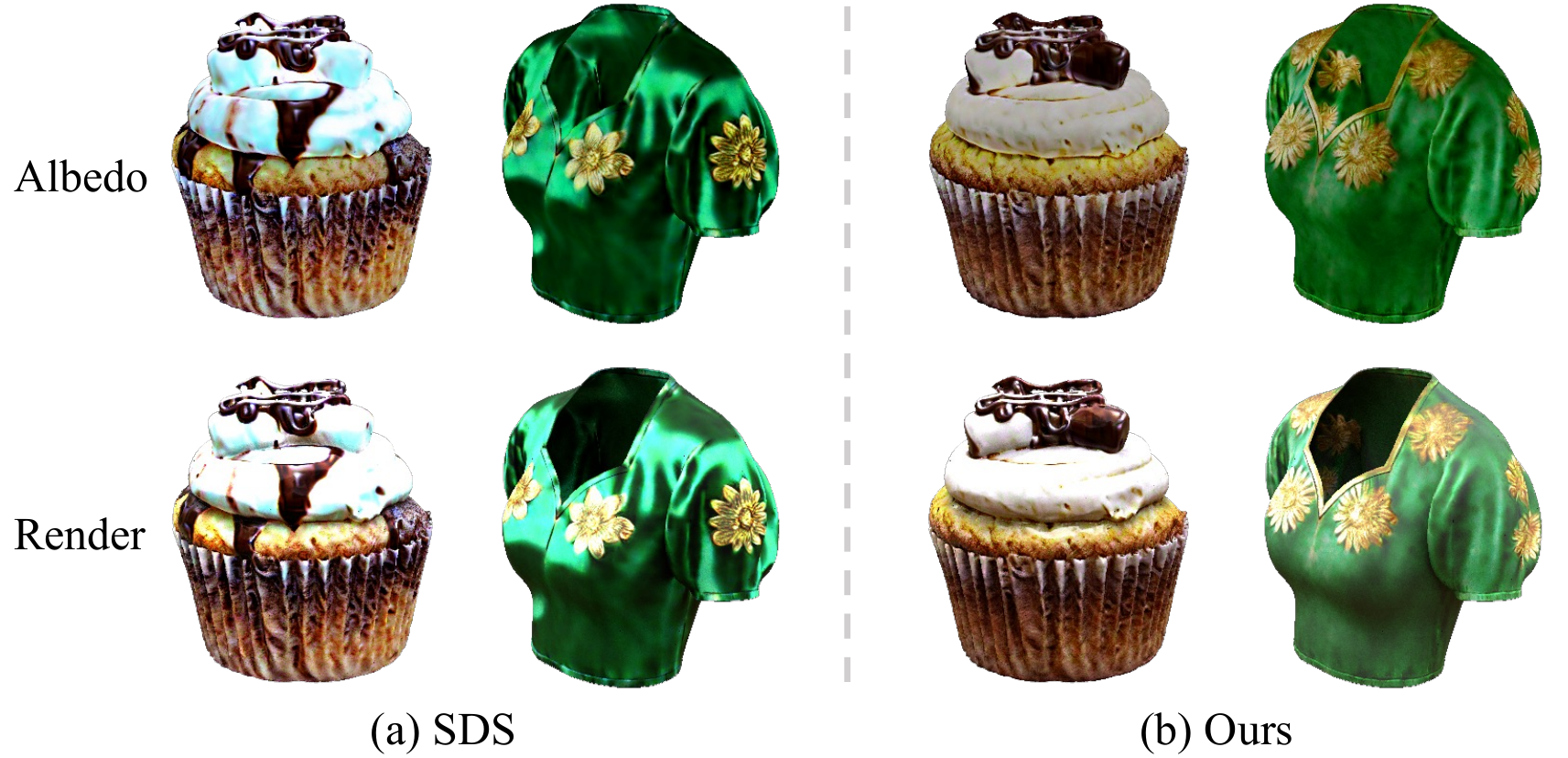}}
  \caption{\revision{Results using our geometry- and light-aware diffusion model with SDS loss (a) and CSD loss (b) for the prompts ``A cupcake with marshmallow and chocolate drizzle toppin'' and ``a green silk blouse with golden flower embroidery''.}}
  \label{fig:sds_ours}
\end{figure}
\subsection{Applying SDS loss on our diffusion model}
\label{sec: sds_appendix}
\revision{Figure~\ref{fig:sds_ours} shows the results of using SDS loss with our geometry- and light-aware diffusion model. Notably, the use of SDS loss results in increased saturation and contrast in the generated appearances, which are less visually appealing. In comparison, our method with CSD loss produces better results. 
}

\subsection{Using text prompts to generate albedo}
\begin{figure}
  \includegraphics[width=\linewidth]{{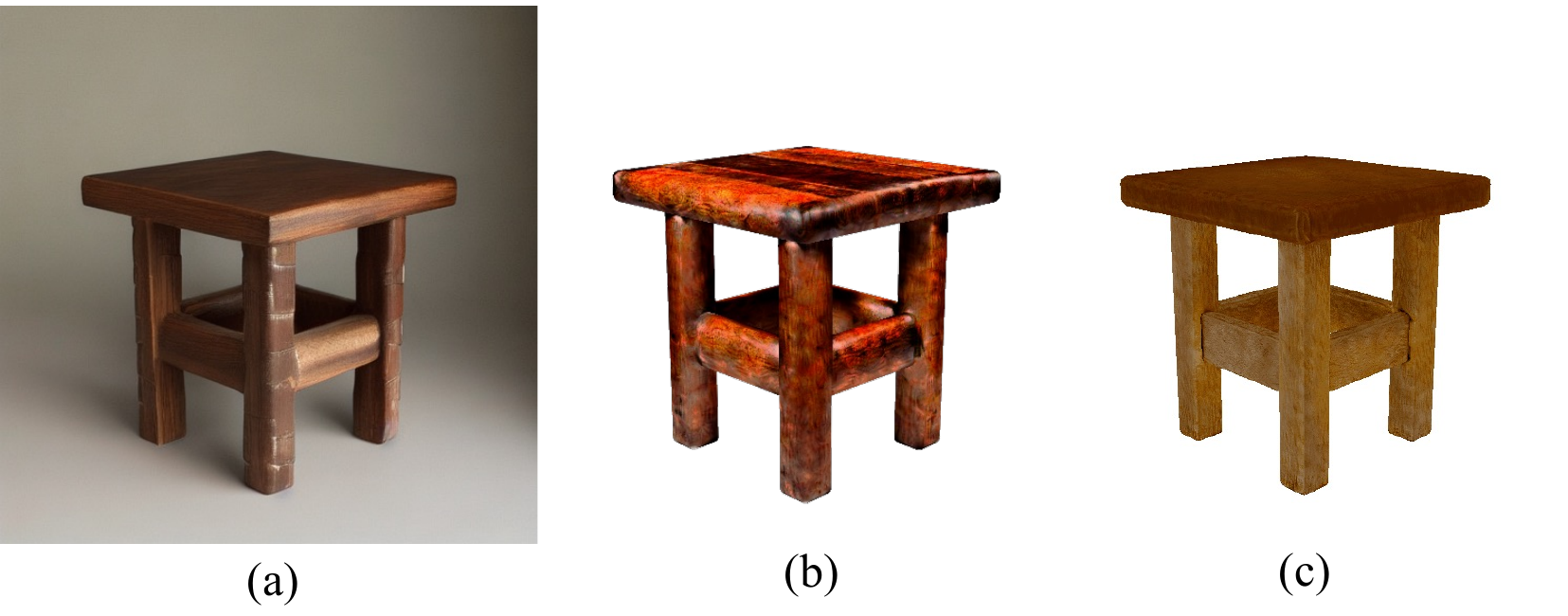}}
  \caption{\revision{(a) An albedo image generated directly by Stable Diffusion using text prompt ``a wooden stool, diffuse albedo map''. (b) An albedo map distilled from a 3D model using CSD loss with the text prompt ``a wooden stool, diffuse albedo map''. (c) An albedo map obtained by our method.}}
  \label{fig:prompt}
\end{figure}

\revision{
Directly using text prompts like ``albedo maps'' in Stable Diffusion cannot correctly generate albedo maps, as illustrated in Fig.~\ref{fig:prompt}. We incorporate the text prompts ``diffuse albedo maps'' to depth-conditioned Stable Diffusion to generate an image as shown in Fig.~\ref{fig:prompt} (a) and also use this text prompt to distill an albedo map as shown in Fig.~\ref{fig:prompt} (b). It is observable that despite the explicit inclusion of ``diffuse albedo'' in the text prompts, the albedo results still contain shading effects. This can be attributed to the inadequate text guidance and a lack of albedo training data within Stable Diffusion.}

\subsection{Different material representations}

\begin{figure}
  \includegraphics[width=\linewidth]{{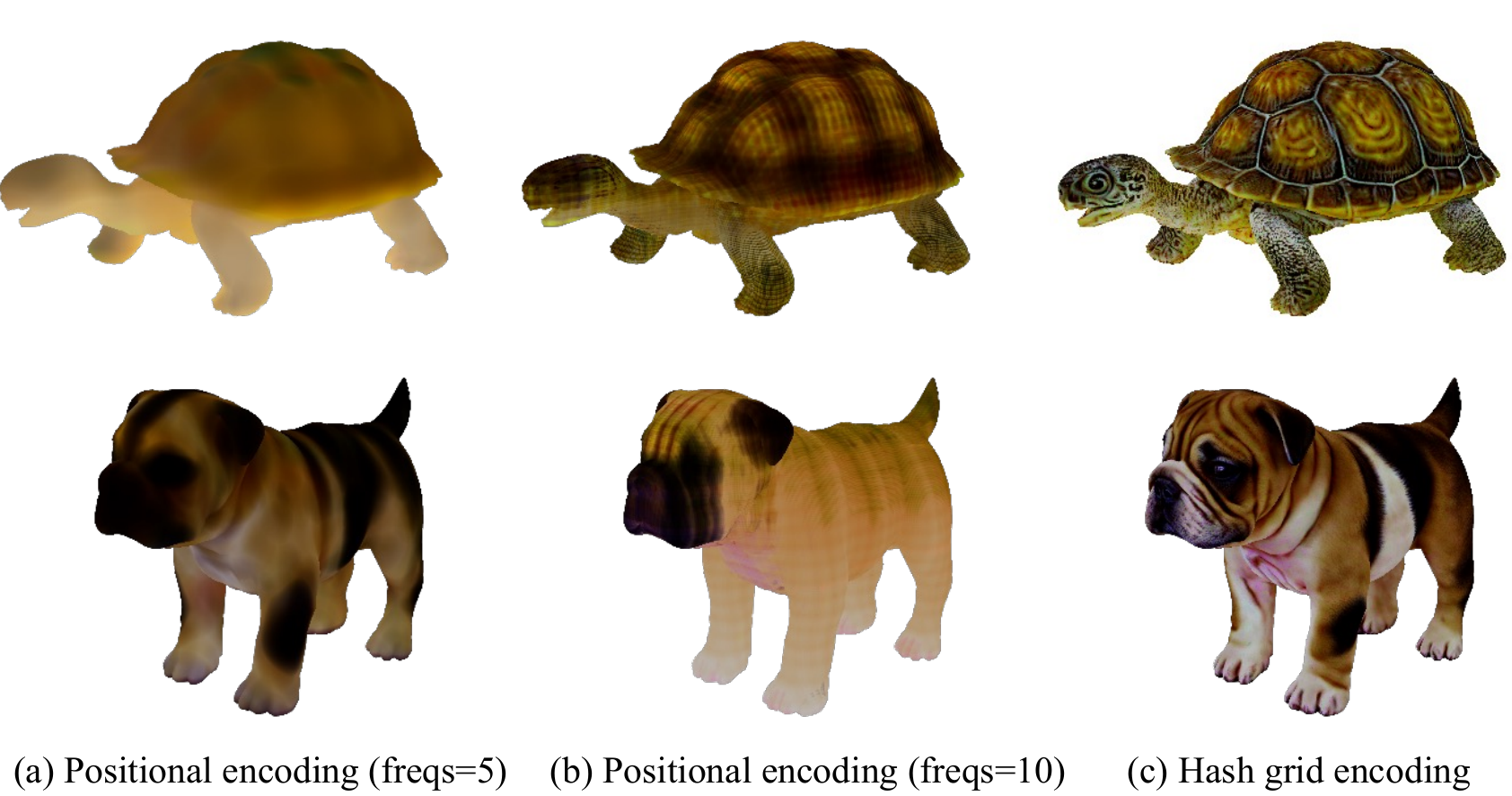}}
  \caption{\revision{The generated albedo maps utilizing positional encoding with different numbers of frequencies (a, b), and hash-grid encoding (c).}}
  \label{fig:hash_grid}
\end{figure}

\revision{In Figure~\ref{fig:hash_grid}, we present the albedo maps achieved through different material representations including Multi-layer Perceptron (MLP) network with positional encoding (PE)~\cite{mildenhall2021nerf} and hash-grid representation~\cite{muller2022instant}. During the distillation process, when employing PE-equipped MLPs, a higher frequency count tends to produce axis-aligned artifacts. Conversely, a lower frequency count fails to adequately capture the fine details of the texture. However, the adoption of a hash-grid-based representation effectively mitigates the aforementioned issues, additionally demonstrating a more rapid convergence rate.}

\subsection{Addition Results}

\begin{figure*}
  \includegraphics[width=\linewidth]{{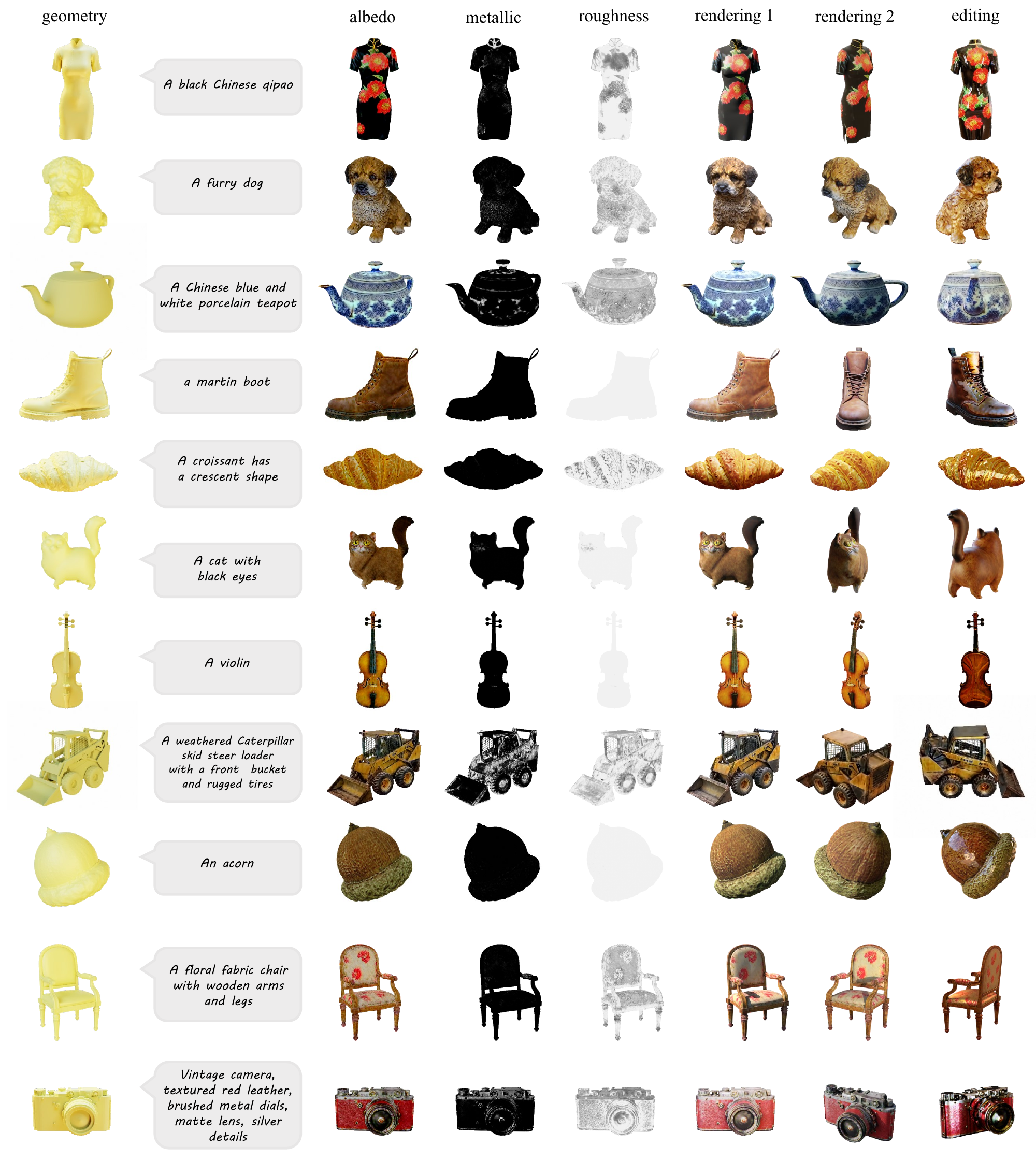}}
  \caption{Generated materials and their renderings and editing results.}
  \label{fig:relight_supp}
\end{figure*}

In Fig. \ref{fig:relight_supp}, we further demonstrate several examples of materials generated based on corresponding text prompts, along with their rendered results under various lighting conditions. We can further perform material editing by adjusting the metallic and roughness values and re-rendering the edited objects.

\subsection{3D Model Attribution}
In this paper, we use 3D models sourced from the Objaverse dataset~\cite{deitke2023objaverse} and Sketchfab~\cite{Sketchfab} under the Creative Commons Attribution 4.0 International (CC BY 4.0) license. The models are utilized without their original textures to focus solely on the impact of our material generation method.

Each model used from Sketchfab is attributed as follows:
\begin{itemize}
\item ``\href{https://sketchfab.com/3d-models/bobcat-machine-7845344823cb4cdcb99963f561e5d866}{Bobcat machine}'' by mohamed ouartassi.
\item ``\href{https://sketchfab.com/3d-models/molino-de-viento---windmill-2ea0a5296d4b49dbad71ce1975c0e3ff}{Molino De Viento \_ Windmill}'' by BC-X.
\item ``\href{https://sketchfab.com/3d-models/medivalhousehouse-for-livingmedivalvilage-ba53607959b0476fb719043c406bc245}{MedivalHouse\ |\ house for living\ |\ MedivalVilage}'' by JFred-chill.
\item ``\href{https://sketchfab.com/3d-models/houseleek-plant-70679a304b324ca8941c214875acf6a9}{Houseleek plant}'' by matousekfoto.
\item \href{https://sketchfab.com/3d-models/jagernaut-beyond-human-977e3a466dbc4c859071e342c6b6151e}{Jagernaut (Beyond Human)} by skartemka.
\item ``\href{https://sketchfab.com/3d-models/grabfigur-fbd44dd62766450abefaa0e43941633e}{Grabfigur}'' by noe-3d.at.
\item ``\href{https://sketchfab.com/3d-models/teenage-mutant-ninja-turtles-raphael-191f64c3a6a44218a98a4d93f44229a9}{Teenage Mutant Ninja Turtles - Raphael}'' by Hellbruch. 
\item ``\href{https://sketchfab.com/3d-models/cat-with-jet-pack-9afc8fd58c0d4f7d827f2007d6ac1e80}{Cat with jet pack}'' by Muru.
\item ``\href{https://sketchfab.com/3d-models/transformers-universe-autobot-showdown-7a3f2d273f354b29b31f247beb62d973}{Transformers Universe: Autobot Showdown}'' by Primus03.
\item ``\href{https://sketchfab.com/3d-models/pigman-f7597d3af7224f7e890710ac27d4d597}{PigMan}'' by Grigorii Ischenko.
\item"\href{https://sketchfab.com/3d-models/bulky-knight-002a90cbf12941b792f9685546a7502c}{Bulky Knight}'' by Arthur Krut.
\item ``\href{https://sketchfab.com/3d-models/sir-frog-chrono-trigger-0af0c15e947143be8fab274841764bf1}{Sir Frog}'' by Adrian Carter.
\item ``\href{https://sketchfab.com/3d-models/infantry-helmet-ba3a571a8077417f80ae0e06150c91d2}{Infantry Helmet}'' by Masonsmith2020.
\item ``\href{https://sketchfab.com/3d-models/sailing-ship-model-ac65e0168e8c423db9c9fdc71397c84e}{Sailing Ship Model}'' by Andrea Spognetta (Spogna). 
\item ``\href{https://sketchfab.com/3d-models/venice-mask-4aace12762ee44cf97d934a6ced12e65}{Venice Mask}'' by DailyArt.
\item ``\href{https://sketchfab.com/3d-models/bouddha-statue-photoscanned-2d71e5b04f184ef89130eb26bc726add}{Bouddha Statue Photoscanned}'' by amcgi.
\item "\href{https://sketchfab.com/3d-models/bunny-c362411a4a744b6bb18ce4ffcf4e7f43}{Bunny}" by vivienne0716.
\end{itemize}

\end{document}


\title{DreamMat: High-quality PBR Materials Generation with Geometry- and Light-aware Stable Diffusion\\
-- \textit{Supplementary Material} --}

\author{Yuqing Zhang}
\authornote{Equal contribution}
\affiliation{
  \institution{State Key Lab of CAD\&CG, Zhejiang University}
  \city{Hangzhou}
  \state{Zhejiang}
  \country{China}
}

\author{Yuan Liu}
\authornotemark[1]
\affiliation{%
  \institution{Tencent Games}
  \city{Shenzhen}
  \country{China}
  }

\author{Zhiyu Xie}
\affiliation{
  \institution{State Key Lab of CAD\&CG, Zhejiang University}
  \city{Hangzhou}
  \state{Zhejiang}
  \country{China}
}

\author{Lei Yang}
\affiliation{%
  \institution{Tencent Games}
  \city{Shenzhen}
  \country{China}
  }
  
\author{Zhongyuan Liu}
\affiliation{%
  \institution{Tencent Games}
  \city{Shenzhen}
  \country{China}
  }
\author{Mengzhou Yang}
\affiliation{%
  \institution{Tencent Games}
  \city{Shenzhen}
  \country{China}
  }
\author{Runze Zhang}
\affiliation{%
  \institution{Tencent Games}
  \city{Shenzhen}
  \country{China}
  }
\author{Qilong Kou}
\affiliation{%
  \institution{Tencent Games}
  \city{Shenzhen}
  \country{China}
  }
\author{Cheng Lin}
\affiliation{%
  \institution{Tencent Games}
  \city{Shenzhen}
  \country{China}
  }
\author{Wenping Wang}
\affiliation{%
  \institution{Texas A\&M University}
  \city{Texas}
  \country{U.S.A}
  }
\author{Xiaogang Jin}
\authornote{Corresponding author.}
\affiliation{
  \institution{State Key Lab of CAD\&CG, Zhejiang University}
  \city{Hangzhou}
  \state{Zhejiang}
  \country{China}
}

\maketitle
\appendix
\section{More Implementation Details}

\subsection{Geometry- and Light-aware diffusion model Training}
\begin{figure*}
  \includegraphics[width=\linewidth]{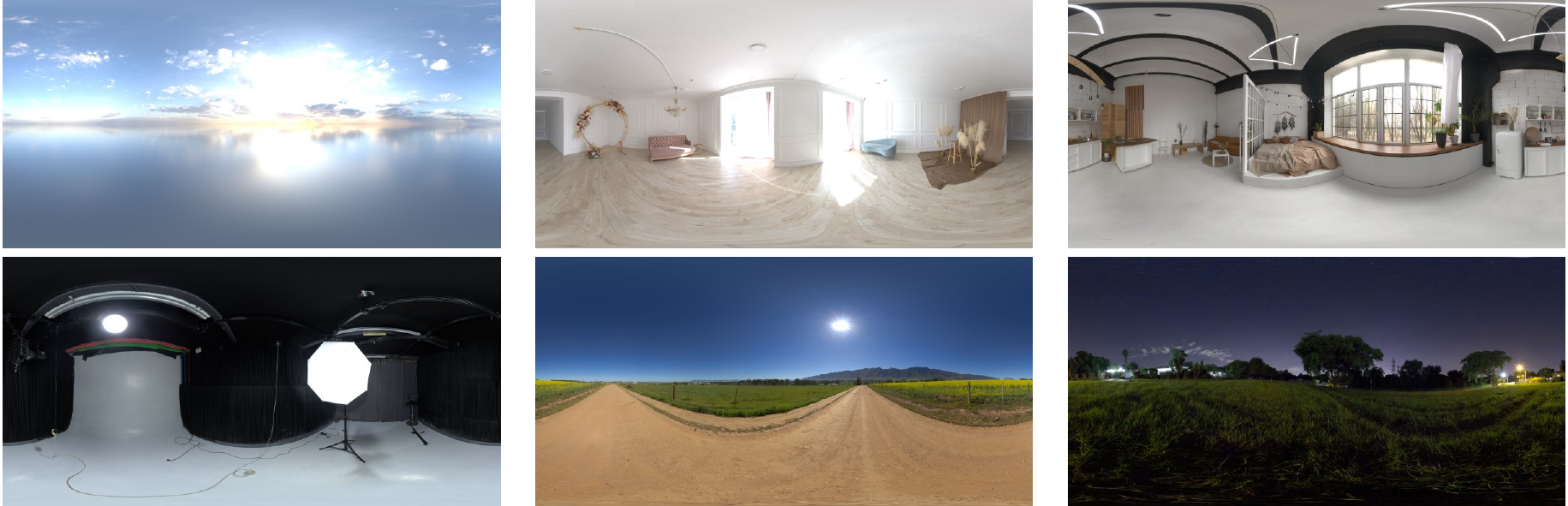}
  \caption{Environment light maps used in geometry- and light-aware diffusion model training and material distillation.}
  \label{fig:environment}
\end{figure*}

The geometry- and light-aware diffusion model is trained using six distinct environment light maps, as illustrated in Fig. \ref{fig:environment}. Within the Objaverse dataset \cite{deitke2023objaverse}, we first exclude objects that failed to meet specific criteria, such as those containing transparent parts, non-mesh structures, or too simple geometry (e.g., a single plane). Then, we randomly choose one environment light map to render both the light condition maps and the final results. During the rendering of the light condition map, we set albedo to white color, metallic to 0.0 and 1.0, and roughness to 0.0, 0.5, and 1.0. 

\begin{figure*}
  \includegraphics[width=\linewidth]{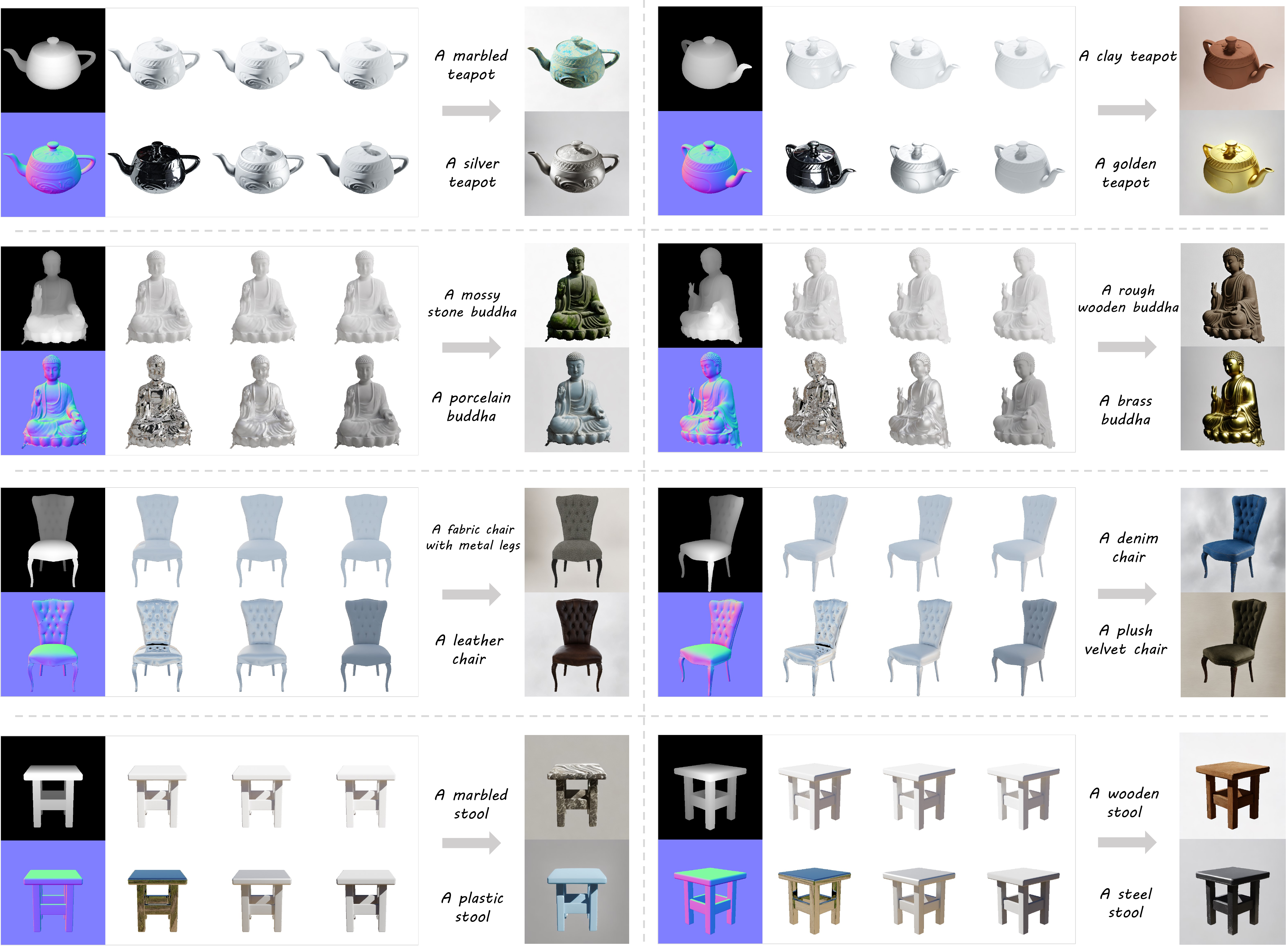}
  \caption{Image generated by the geometry- and light-aware diffusion model with different text prompts.}
  \label{fig:controlnet}
\end{figure*}

Our diffusion model uses a ControlNet~\cite{zhang2023adding} with a condition map of 22 channels. The learning rate is set as 1e-5 with a single-step gradient accumulation and a batch size of 256. The model is trained for a total of 3 epochs, which uses 8 V100 GPUs for 3 days. To demonstrate the versatility and quality of the ControlNet, Fig. \ref{fig:controlnet} illustrates the generated results using different text prompts under different lighting conditions.


\subsection{Baseline implementation details}
For TANGO~\cite{chen2022tango}, we follow the official implementation and set the learning rate to 5e-4, which is decayed by 0.7 in every 500 iterations. We iterate 1500-3000 times for each object until convergence. Since TANGO uses position encoding $\beta(l) =  [\cos(2\pi Bl), \sin(2\pi Bl) ]^T$ to provide high-frequency details of the generated materials, where B is a random Gaussian matrix whose entry is randomly drawn from $N (0, \sigma^2)$ (different examples in the open source code use different position encoding parameters), we find that the final result is greatly influenced by the PE parameters. Therefore, we carefully tune the $\sigma$ and frequency number for each example to achieve the best performance. 

For TEXTure~\cite{richardson2023texture} and Text2Tex~\cite{chen2023text2tex}, we use their original implementations to generate texture results, and then apply NVdiffrec~\cite{Munkberg_2022_CVPR} for subsequent material decomposition. In the material decomposition stage, we fix the geometry and lighting with a learning rate of 0.01 for 3000 iterations.

For Fantasia3D~\cite{Chen_2023_ICCV}, we only use its second stage to generate the objects’ materials. In this stage, we fix the environmental lighting and geometry, optimizing the albedo, roughness, and metallic with prompts related to the viewpoint for 3000 iterations. The learning rate is set to 0.01.
\begin{figure*}
\includegraphics[width=0.95\linewidth]{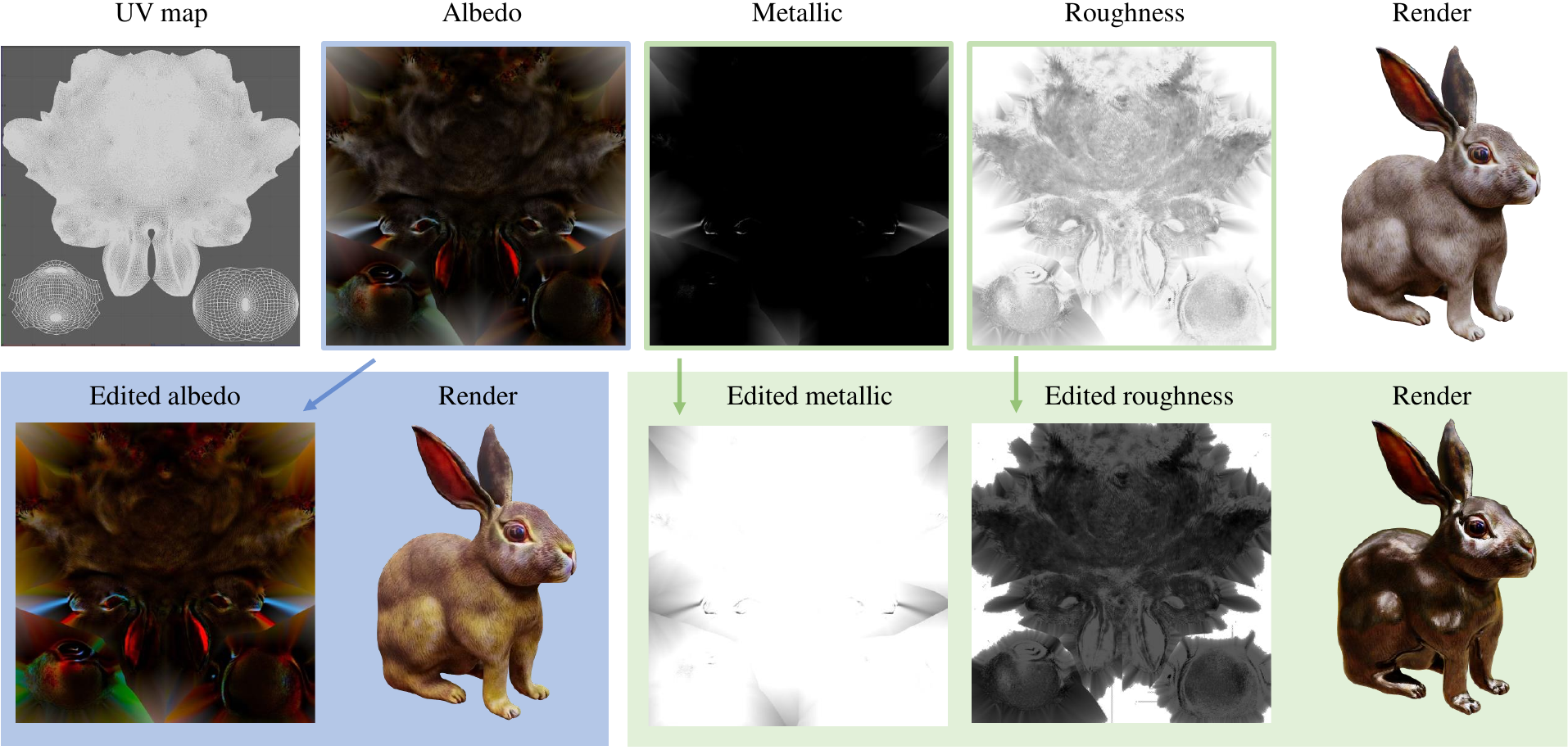}
  \caption{\revision{\textbf{Texture maps and material editing.} Top: Generated material maps utilizing DreamMat and the rendering results. Bottom: Edited material maps using 2D image editing techniques and their rendering results.}}
  \label{fig:uv}
\end{figure*}
\subsection{UV Mapping and Material Editing}
During the texture map output phase, we employ UV mapping to sample the generated appearance. The model's inherent UV map can be utilized, or alternatively, a UV map can be automatically generated using xatlas~\cite{xatlas}. Following the previous methodologies~\cite{Munkberg_2022_CVPR, Chen_2023_ICCV}, we apply the UV edge padding technique to extend the boundaries of UV islands and fill in empty regions. The output texture maps are shown in Fig.~\ref{fig:uv}, which can be seamlessly integrated into graphics engines.

Furthermore, these texture maps can be imported into various image editing software (e.g., Photoshop) for material editing. As demonstrated in the bottom row of Fig.~\ref{fig:uv}, we adjust the albedo's saturation, invert the metallic properties, and modify the overall brightness of the roughness. These adjustments enable the achievement of diverse rendering outcomes under the same lighting conditions.

\begin{figure}
  \includegraphics[width=\linewidth]{{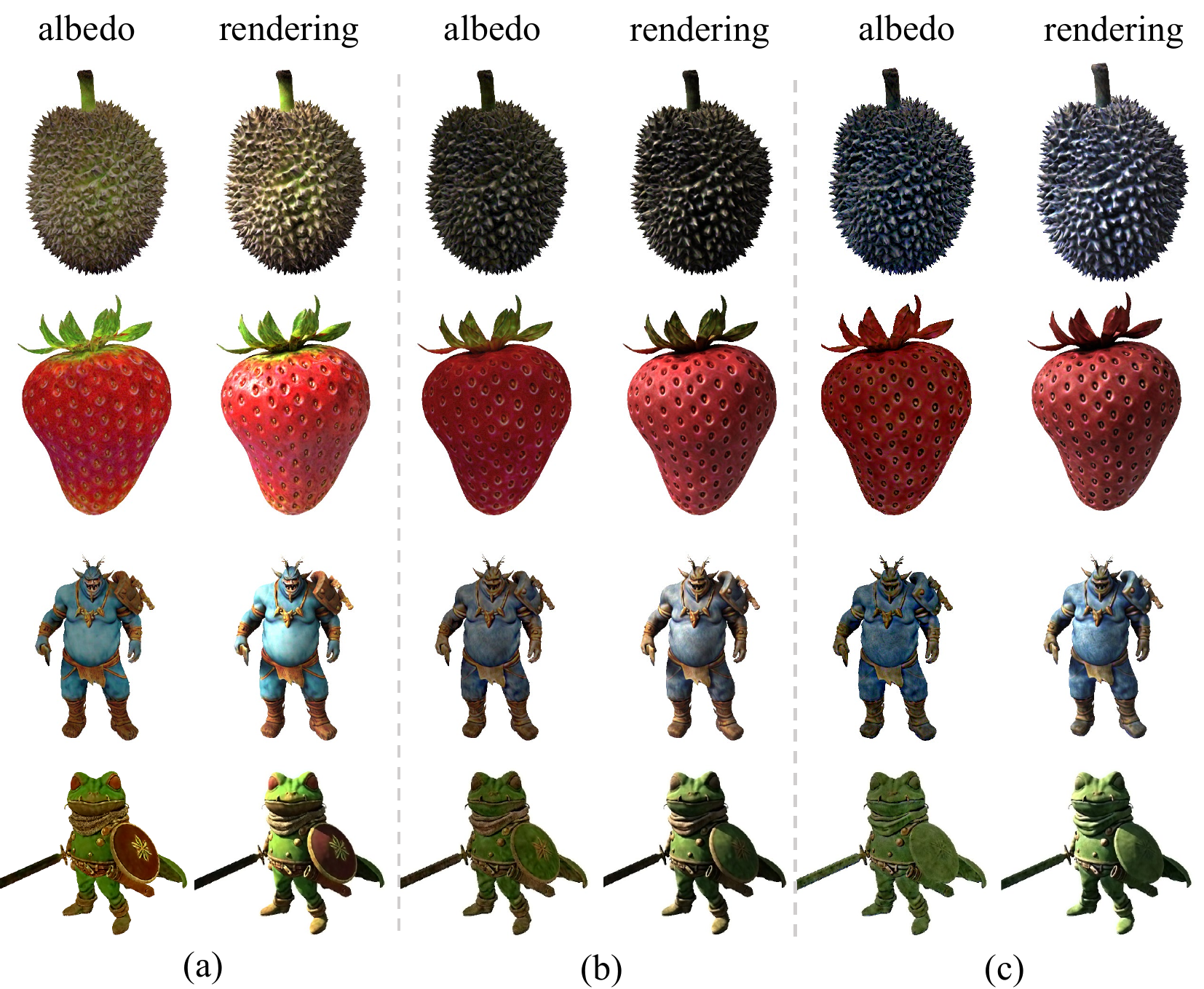}}
  \caption{(a) Our result, incorporating condition maps concurrently to $\epsilon_{\phi}(I_t; y_{\text{pos}}, t)$ ,  $ \epsilon_{\phi}(I_t; t) $, $\epsilon_{\phi}(I_t; y_{\text{neg}}, t)$; (b) The result when condition maps are only added to $ \epsilon_{\phi}(I_t; y_{\text{pos}}, t)$, $\epsilon_{\phi}(I_t; y_{\text{neg}}, t)$; (c) The result with condition maps only added into $\epsilon_{\phi}(I_t; y_{\text{pos}}, t)$}
  \label{fig:csd_supp}
\end{figure}

\section{Condition Maps to CSD Loss}
A proper control strength would be important in the CSD distillation process. As written in Sec 3.2, $\delta(I_t)$ consists of three components: $ \epsilon_{\phi}(I_t; y_{\text{pos}}, t)$, $ \epsilon_{\phi}(I_t; t) $, and $\epsilon_{\phi}(I_t; y_{\text{neg}}, t)$, representing the scenarios with positive prompts, without prompts, and with negative prompts, respectively. Following the implementation in the CSD~\cite{yu2023csd} source code, we applied the condition map to all three components, resulting in the outcomes shown in Fig.\ref{fig:csd_supp} (a). Additionally, we experimented with applying the condition map only to $ \epsilon_{\phi}(I_t; y_{\text{pos}}, t)$ and $\epsilon_{\phi}(I_t; y_{\text{neg}}, t)$ and found that similar results could be achieved, but with some color distortion, as shown in Fig. \ref{fig:csd_supp}(b). Moreover, applying the condition map solely to $ \epsilon_{\phi}(I_t; y_{\text{pos}}, t)$ suffers from a similar problem of tonal distortion.

\section{Entanglement of Materials and Lighting}

\begin{figure}
  \includegraphics[width=\linewidth]{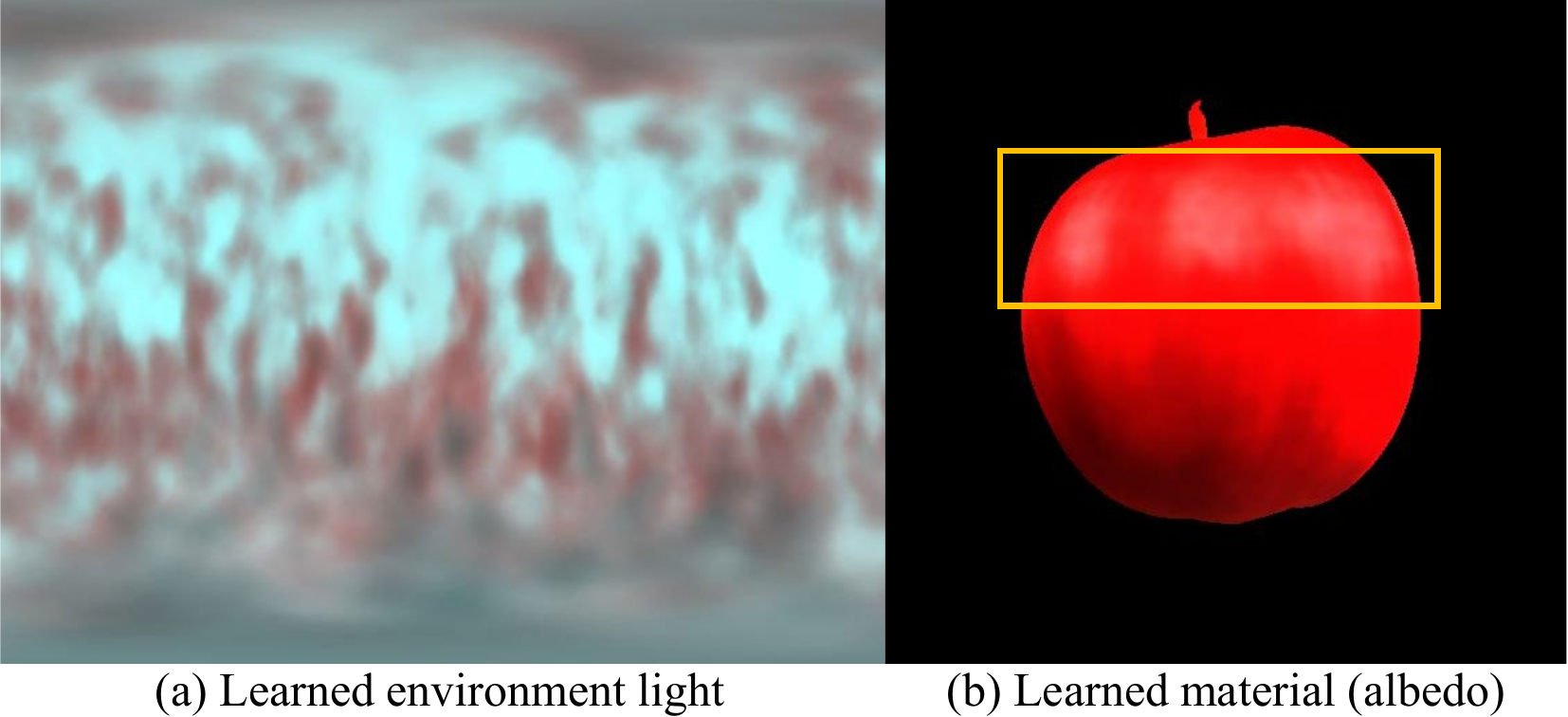}
  \caption{\textbf{Illustration of Material-Lighting Entanglement in Optimized Renderings.} (a) the environment lighting map derived from the SDS optimization process without constraints, where material attributes of the apple are inadvertently encoded. (b) the albedo of the apple, which exhibits shading effects including highlights and shadows, indicative of the entanglement with the environmental lighting. This visual evidence supports the need for employing known light maps to mitigate the ill-posed challenges of inverse rendering.}
  \label{fig:entanglement}
\end{figure}

In Fig.~\ref{fig:entanglement}, we show the results of directly optimizing both lighting and materials using SDS losses, without imposing any constraints on the diffusion process or the materials themselves. For lighting representation, we adopt the model used in NeRO~\cite{liu2023nero}, outputting the results as an environment lighting map after sampling. Notably, we observe that the learned environmental lighting contains substantial information about the material properties of the apple, while the apple's albedo retains numerous shading effects, such as highlights and shadows. This phenomenon stems from the ill-posed nature of inverse rendering, a challenge that becomes more pronounced when the lighting in stable diffusion-generated images is arbitrary. Consequently, the utilization of known environmental light maps is crucial to address these issues.



\section{Addition Results}

\begin{figure*}
  \includegraphics[width=\linewidth]{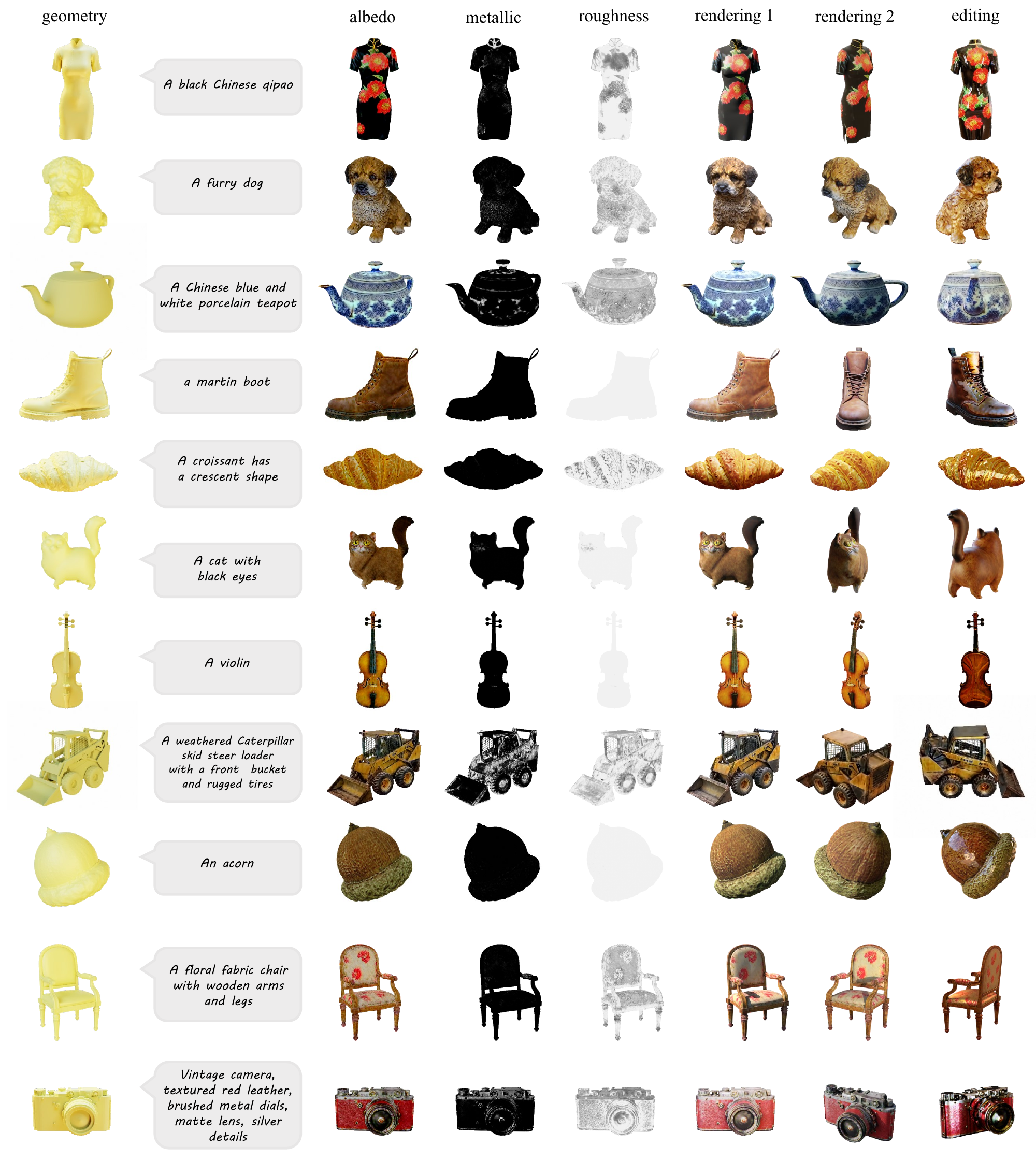}
  \caption{Generated materials and their renderings and editing results.}
  \label{fig:relight_supp}
\end{figure*}

In Fig. \ref{fig:relight_supp}, we further demonstrate several examples of materials generated based on corresponding text prompts, along with their rendered results under various lighting conditions. We can further perform material editing by adjusting the metallic and roughness values and re-rendering the edited objects.

\begin{figure*}
  \includegraphics[width=\linewidth]{{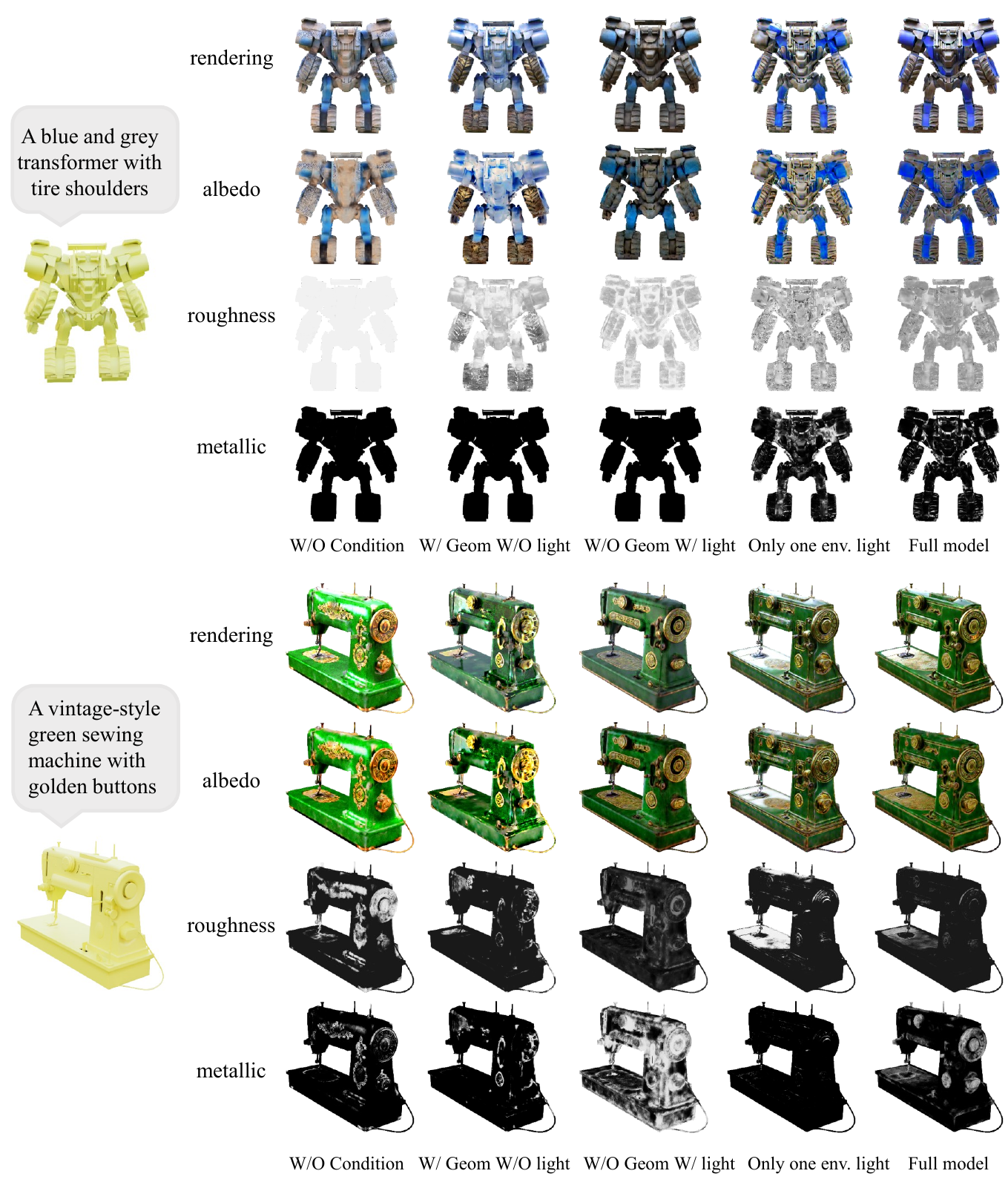}}
  \caption{Ablation results.}
  \label{fig:ablation_supp}
\end{figure*}

In Fig. \ref{fig:ablation_supp}, additional ablation results on the `Transformer' and `sewing machine' cases are illustrated, further demonstrating the effectiveness of each component in our method. 

\begin{figure*}
  \includegraphics[width=0.92\linewidth]{{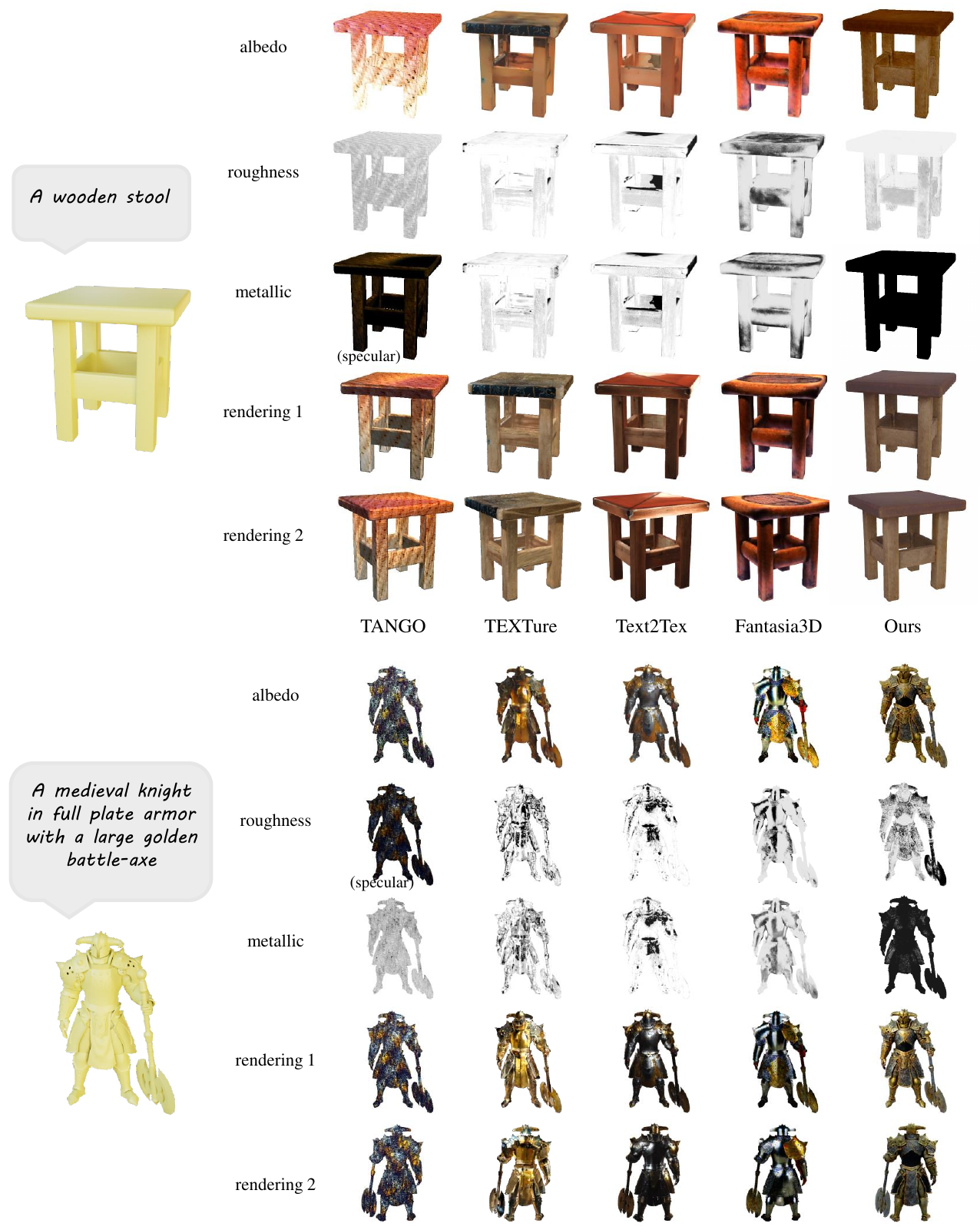}}
  \caption{Additional qualitative comparison with TANGO~\cite{chen2022tango}, TEXTure~\cite{yu2023texture}, Text2Tex~\cite{yu2023text} and Fantasia3D~\cite{Chen_2023_ICCV}.}
  \label{fig:comparison_supp}
\end{figure*}

More qualitative comparisons are presented in Fig. \ref{fig:comparison_supp}. TANGO decomposes the specular F0 term, which represents the base reflectivity of the surface, slightly differing in definition from the metallic maps of the other four methods. In the case of “stool”, due to self-occlusion from the viewpoint, other methods fail to correctly handle the light and shadow relationship, resulting in the separation of shadows from the albedo. Moreover, for wooden materials, the metallic map should be closer to 0. For complex objects like “knight”, we find that other methods are unable to generate high-quality texture details, and there is a lot of highlight residue left in the albedo map. TEXTure and Text2Tex often have issues with blurriness or geometric inconsistencies, while TANGO exhibits obvious grid-like artifacts. The results from Fantasia3D contain a lot of lighting information in the albedo, as well as issues with oversaturation and lack of detail.
\section{3D Model Attribution}
In this paper, we use 3D models sourced from the Objaverse dataset~\cite{deitke2023objaverse} and Sketchfab~\cite{Sketchfab} under the Creative Commons Attribution 4.0 International (CC BY 4.0) license. The models are utilized without their original textures to focus solely on the impact of our material generation method.

Each model used from Sketchfab is attributed as follows:
\begin{itemize}
\item "\href{https://sketchfab.com/3d-models/bobcat-machine-7845344823cb4cdcb99963f561e5d866}{Bobcat machine}" by mohamed ouartassi.
\item "\href{https://sketchfab.com/3d-models/molino-de-viento---windmill-2ea0a5296d4b49dbad71ce1975c0e3ff}{Molino De Viento \_ Windmill}" by BC-X.
\item "\href{https://sketchfab.com/3d-models/medivalhousehouse-for-livingmedivalvilage-ba53607959b0476fb719043c406bc245}{MedivalHouse\ |\ house for living\ |\ MedivalVilage}" by JFred-chill.
\item "\href{https://sketchfab.com/3d-models/houseleek-plant-70679a304b324ca8941c214875acf6a9}{Houseleek plant}" by matousekfoto.
\item \href{https://sketchfab.com/3d-models/jagernaut-beyond-human-977e3a466dbc4c859071e342c6b6151e}{Jagernaut (Beyond Human)} by skartemka.
\item "\href{https://sketchfab.com/3d-models/grabfigur-fbd44dd62766450abefaa0e43941633e}{Grabfigur}" by noe-3d.at.
\item "\href{https://sketchfab.com/3d-models/teenage-mutant-ninja-turtles-raphael-191f64c3a6a44218a98a4d93f44229a9}{Teenage Mutant Ninja Turtles - Raphael}" by Hellbruch. 
\item "\href{https://sketchfab.com/3d-models/cat-with-jet-pack-9afc8fd58c0d4f7d827f2007d6ac1e80}{Cat with jet pack}" by Muru.
\item "\href{https://sketchfab.com/3d-models/transformers-universe-autobot-showdown-7a3f2d273f354b29b31f247beb62d973}{Transformers Universe: Autobot Showdown}" by Primus03.
\item "\href{https://sketchfab.com/3d-models/pigman-f7597d3af7224f7e890710ac27d4d597}{PigMan}" by Grigorii Ischenko.
\item"\href{https://sketchfab.com/3d-models/bulky-knight-002a90cbf12941b792f9685546a7502c}{Bulky Knight}" by Arthur Krut.
\item "\href{https://sketchfab.com/3d-models/sir-frog-chrono-trigger-0af0c15e947143be8fab274841764bf1}{Sir Frog}" by Adrian Carter.
\item "\href{https://sketchfab.com/3d-models/infantry-helmet-ba3a571a8077417f80ae0e06150c91d2}{Infantry Helmet}" by Masonsmith2020.
\item "\href{https://sketchfab.com/3d-models/sailing-ship-model-ac65e0168e8c423db9c9fdc71397c84e}{Sailing Ship Model}" by Andrea Spognetta (Spogna). 
\item "\href{https://sketchfab.com/3d-models/venice-mask-4aace12762ee44cf97d934a6ced12e65}{Venice Mask}" by DailyArt.
\item "\href{https://sketchfab.com/3d-models/bouddha-statue-photoscanned-2d71e5b04f184ef89130eb26bc726add}{Bouddha Statue Photoscanned}" by amcgi.
\item "\href{https://sketchfab.com/3d-models/bunny-c362411a4a744b6bb18ce4ffcf4e7f43}{Bunny}" by vivienne0716.
\end{itemize}

\section{User Study Examples}
In our user study, the participants were comprised of research professionals and senior art practitioners from related fields, all of whom possessed an in-depth understanding of PBR materials. We give a comprehensive presentation of the 20 examples used in the user study, along with their corresponding questions, in the final pages of these supplementary materials.

\begin{figure*}
  \includegraphics[width=0.9\linewidth]{{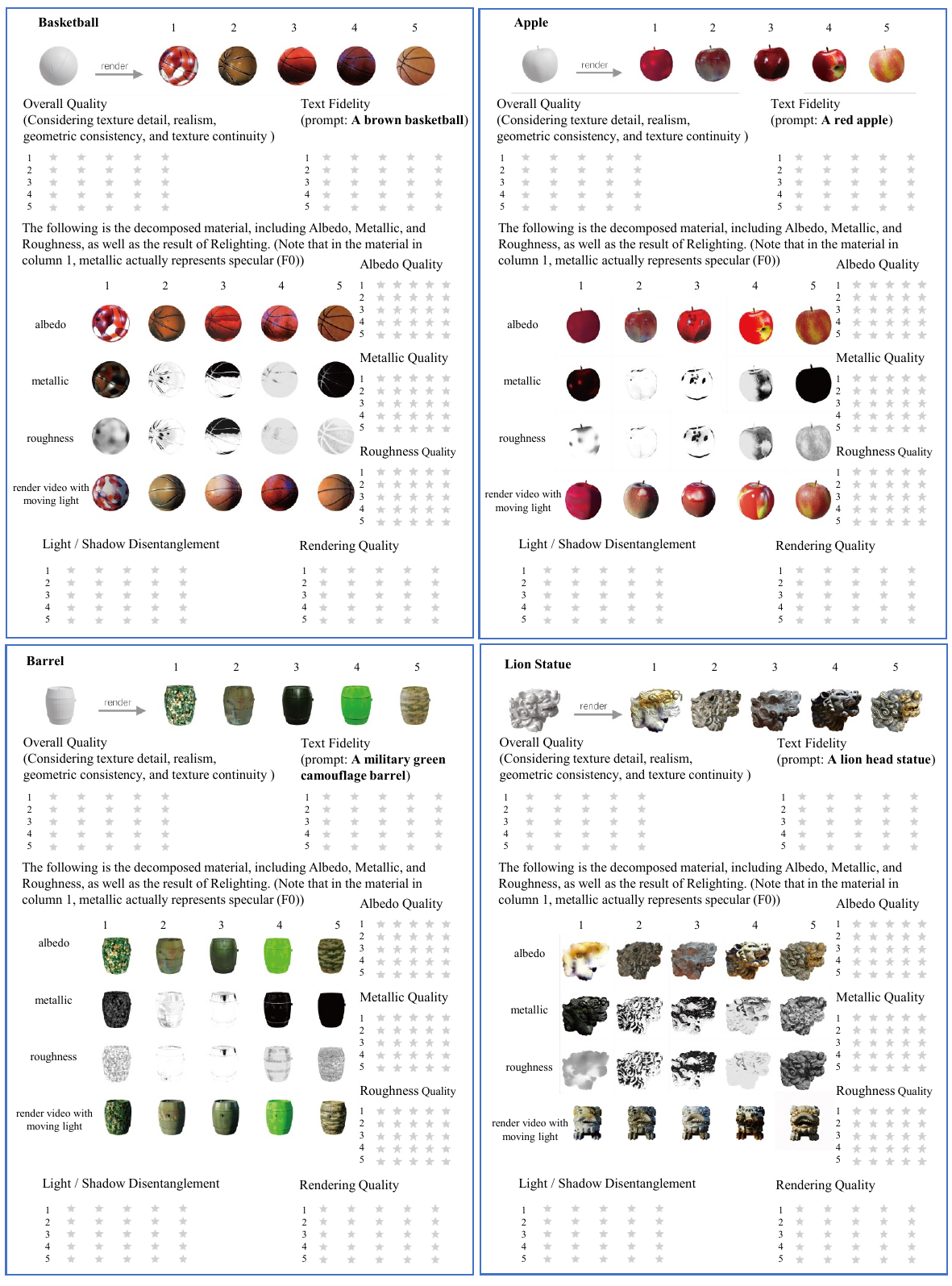}}
  \label{fig:user_study_1}
\end{figure*}

\begin{figure*}
  \includegraphics[width=0.9\linewidth]{{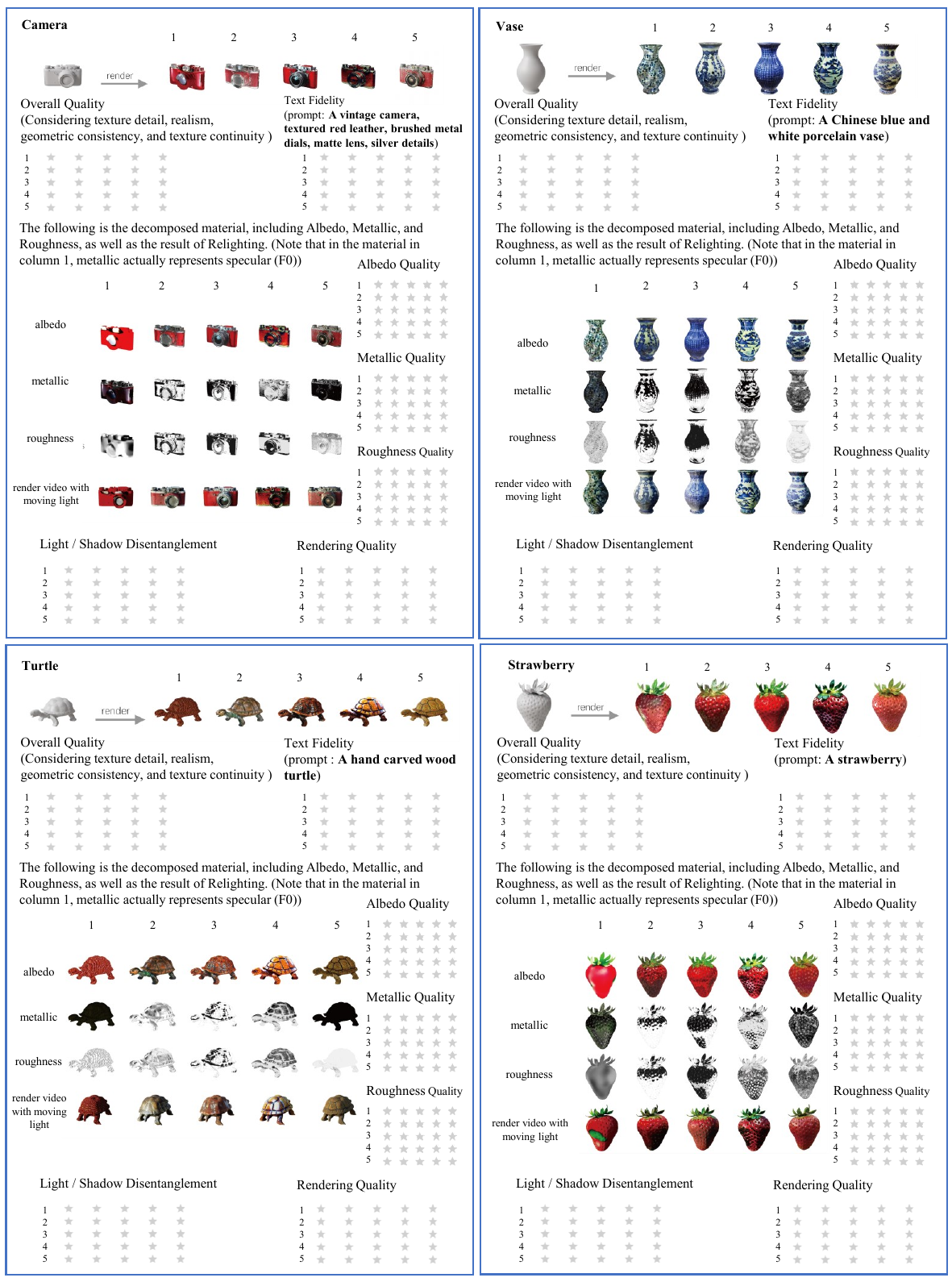}}
  \label{fig:user_study_2}
\end{figure*}

\begin{figure*}
  \includegraphics[width=0.9\linewidth]{{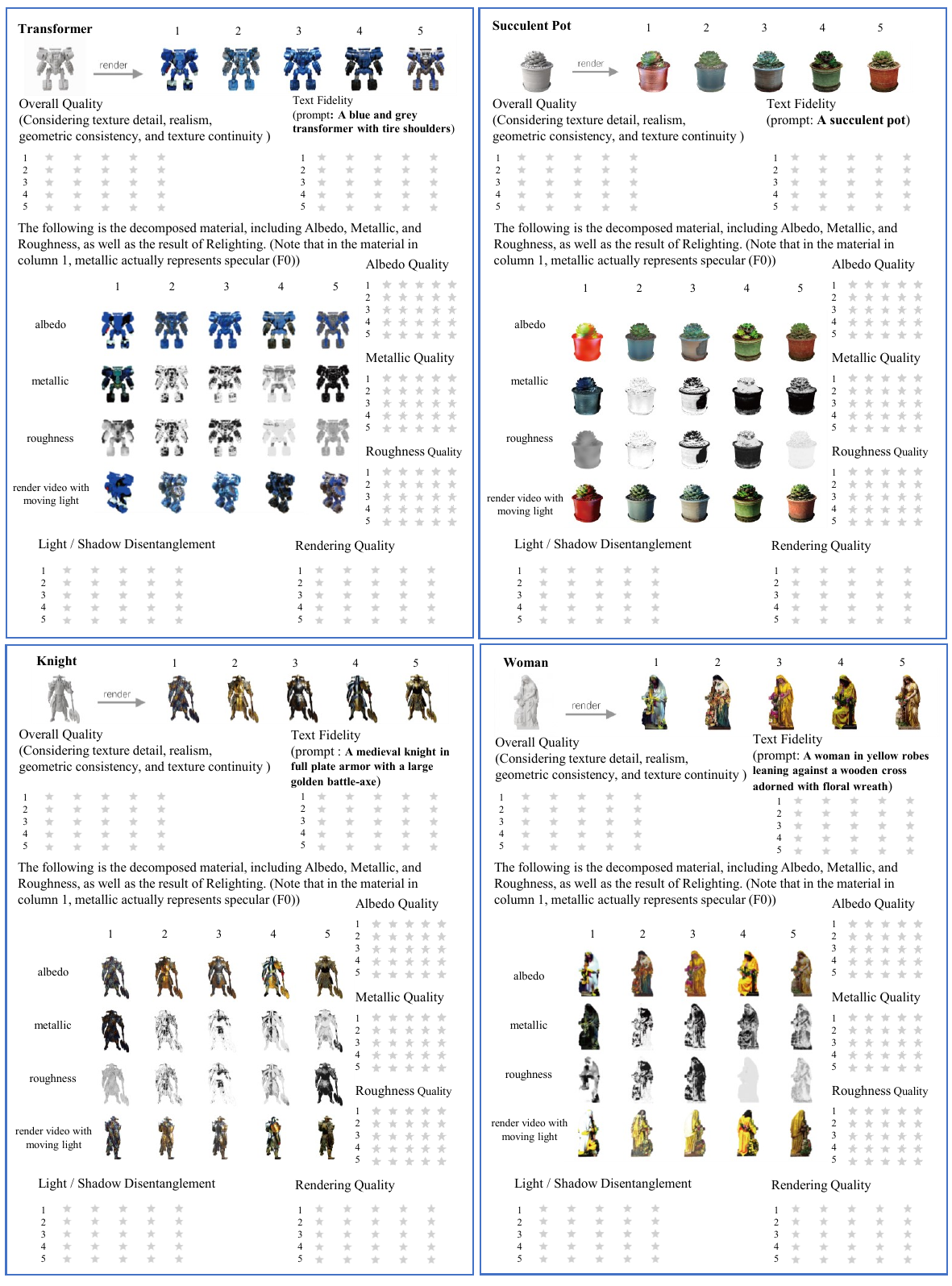}}
  \label{fig:user_study_3}
\end{figure*}

\begin{figure*}
  \includegraphics[width=0.9\linewidth]{{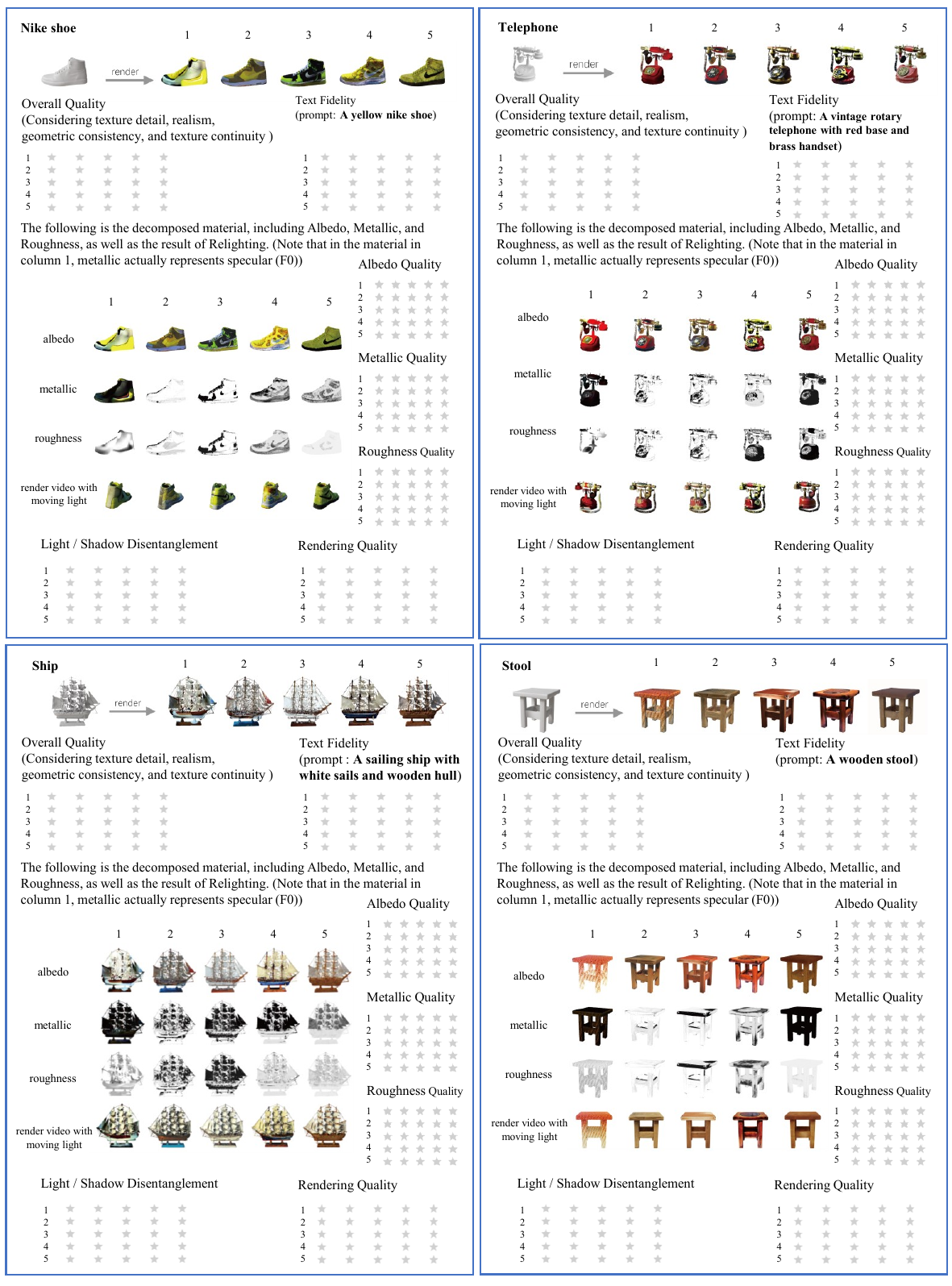}}
  \label{fig:user_study_4}
\end{figure*}

\begin{figure*}
  \includegraphics[width=0.9\linewidth]{{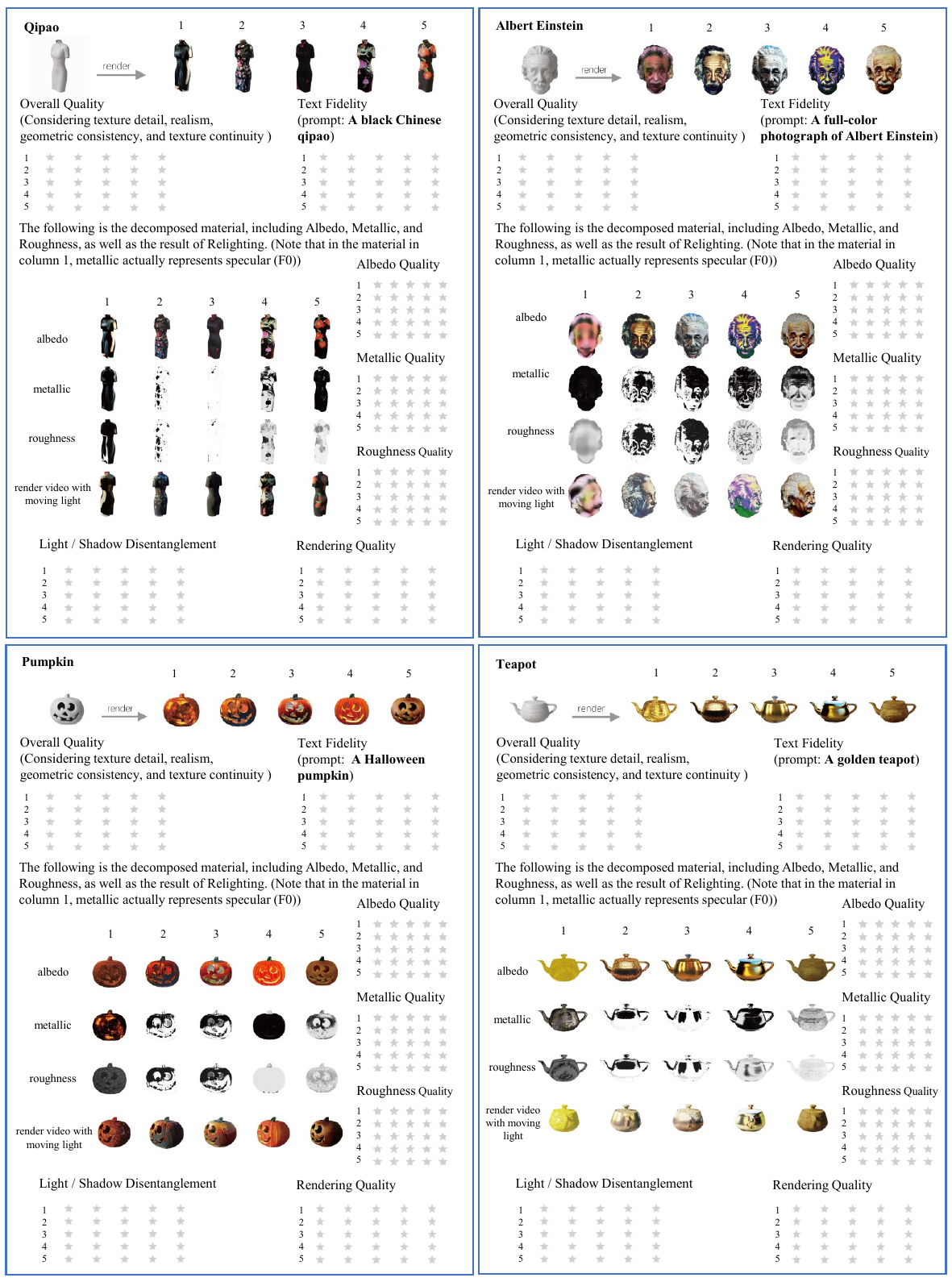}}
  \label{fig:user_study_5}
\end{figure*}


\bibliographystyle{ACM-Reference-Format}
\bibliography{ref}
